\documentclass[manuscript,screen]{acmart}

\usepackage{xspace}
\usepackage{booktabs}
\usepackage{url}
\usepackage{color,soul}
\usepackage{graphicx}
\usepackage{multirow}
\usepackage{amsmath}
\usepackage{array}
\usepackage{tabularx}
\usepackage{longtable}
\usepackage{caption}
\usepackage{subcaption}
\usepackage{tcolorbox}
\usepackage{adjustbox}
\usepackage{xcolor, colortbl}
\usepackage{framed}
\usepackage{natbib}
\usepackage{xcolor}

\graphicspath{{tex/img/}}

\definecolor{light-gray}{gray}{0.95}

\newenvironment{conditions*}
  {\par\vspace{\abovedisplayskip}\noindent
   \tabularx{\columnwidth}{>{$}l<{$} @{${}={}$} >{\raggedright\arraybackslash}X}}
  {\endtabularx\par\vspace{\belowdisplayskip}}

\newcommand{\rust}{\textsf{Rust}\xspace}
\newcommand{\stkovflw}{\textsf{Stack Overflow}\xspace}
\newcommand{\rustforum}{\textsf{Rust Forum}\xspace}

\newcommand{\rqone}{How the prevalence of KUs differs between the concrete information needs of developers and the \rust documentation?}
\newcommand{\rqtwo}{How many KUs are absent, divergent, or convergent in the \rust documentation?}
\newcommand{\rqthree}{Do convergent, divergent, and absent KUs exhibit different awareness in Q\&A websites?}

\newlength{\freewidth}
\AtBeginDocument{%
  \providecommand\BibTeX{{%
    \normalfont B\kern-0.5em{\scshape i\kern-0.25em b}\kern-0.8em\TeX}}}

\setcopyright{acmcopyright}
\copyrightyear{2021}
\acmYear{2021}
\acmDOI{10.1145/1122445.1122456}

\acmConference[TOSEM]{Transactions on Software Engineering and Methodology}{October 2021}{}
\acmBooktitle{ACM Transactions on Software Engineering and Methodology}

\begin{document}

\title[Assessing the alignment between the information needs of developers and Rust documentation]{Assessing the alignment between the information needs of developers and the documentation of programming languages: A case study on Rust}

\author{Filipe R. Cogo}
\email{filipe.roseiro.cogo1@huawei.com}
\affiliation{%
  \institution{Centre for Software Excellence, Huawei}
  \streetaddress{275 Queen St.}
  \city{Kingston}
  \state{Ontario}
  \country{Canada}
  \postcode{K7K1B7}
}


\author{Xin Xia}
\affiliation{%
  \institution{Software Engineering Application Technology Lab, Huawei}
  \city{Shenzhen}
  \country{China}}
\email{xin.xia@huawei.com}

\author{Ahmed E. Hassan}
\affiliation{%
  \institution{School of Computing, Queen's University}
  \city{Kingston}
  \country{Canada}}
\email{ahmed@cs.queensu.ca}

\renewcommand{\shortauthors}{Cogo, et al.}

\begin{abstract}
Programming language documentation refers to the set of technical documents that provide application developers with a description of the high-level concepts of a language (e.g., manuals, tutorials, and API references). Such documentation is essential to support application developers in the effective use of a programming language. One of the challenges faced by documenters (i.e., personnel that design and produce documentation for a programming language) is to ensure that documentation has relevant information that aligns with the concrete needs of developers. In this paper, we present an automated approach to support documenters in evaluating the differences and similarities between the concrete information need of developers and the current state of documentation (a problem that we refer to as the topical alignment of a programming language documentation). Our approach leverages semi-supervised topic modelling that uses domain knowledge to guide the derivation of topics. We initially train a baseline topic model from a set of \rust-related Q\&A posts that represent the concrete information needs of developers. We then use this baseline model to determine the distribution of topic probabilities of each document of the official \rust documentation. Afterwards, we assess the similarities and differences between the topics of the Q\&A posts and the official documentation. Our results show that there is a relatively high level of topical alignment in \rust documentation. Still, information about specific topics is scarce in both the Q\&A websites and the documentation, particularly related topics with programming niches such as network, game, and database development. For other topics (e.g., related topics with language features such as structs, patterns and matchings, and foreign function interface), information is only available on Q\&A websites while lacking in the official documentation. Finally, we discuss implications for programming language documenters, particularly how to leverage our approach to prioritize topics that should be added to the documentation.
\end{abstract}



\maketitle

\section{Introduction}
\label{sec:introduction}

\emph{Programming language documentation} is the set of technical documents that describe high-level concepts of a programming language and allow application developers to learn and use the language in an effective way~\cite{Yngve:communacm_pl_doc:1963}. This documentation is officially maintained and distributed and typically includes a description of the language concepts, reference manuals, usage examples, and tooling tutorials. Examples of such a documentation include the \textsf{Java Documentation}\footnote{\url{https://docs.oracle.com/en/java/}}, the \textsf{Rust Documentation}\footnote{\url{https://www.rust-lang.org/learn}}, and the \textsf{Go Documentation}\footnote{\url{https://golang.org/doc/}}, among others. Without a sufficient set of documents, both novice and experienced developers lack the necessary information about using a programming language to address their needs and improve their knowledge. 

The documentation of a programming language needs to be periodically updated and to follow along with the language evolution. To produce informative documents and to plan the allocation of resources to documentation related activities, documenters (i.e., personnel devoted to writing documentation) should be able to identify the most important and practical topics that developers need and to compare such topics against the current version of the documentation~\cite{robillard:on_demand_doc:2017}. Documenters should also be able to identify topics of the documentation that are not well covered and prioritize topics that should be improved. Similarly, documenters should be able to identify sufficiently documented topics (in relation to developers' information needs) to avoid redundancies in the maintained documentation. We refer to \emph{topical alignment} of the documentation of a programming language as the difference between the topics that make up the concrete information needs of developers and the covered topics by the current state of the documentation. Maximizing the topical alignment of the produced documentation is not a trivial task, as documenters are not always aware of the concrete information needs of developers.

Complete and accurate documentation is of fundamental importance to provide developers with information about the usage of a programming language. Despite its importance, assessing the topical alignment of the documentation of a programming language is an open research problem~\cite{robillard:on_demand_doc:2017, treude:extracting_dev_tasks:2015}. Typically, communities and sponsors around different programming languages employ expensive and time-consuming survey methods to assess the topical alignment of their documentation~\cite{java:documentation_survey:2007,rust:documentation_survey:2020,go:documentation_survey:2020}. In this paper, we describe a machine learning-based approach to tackle this important problem. The main idea behind our approach is to build two models -- one that represents the information needs of developers and another one that represents the covered topics by the current documentation of the language -- and then assess the similarity and dissimilarities between the models. To build such models, we use semi-supervised topic modelling~\cite{gallager:topiccorex:2017} that leverages domain knowledge, represented by a set of associated keywords with each topic (a.k.a. \emph{anchor words}), to guide the derivation of latent topics from textual information~\cite{andrzejewski:icml:2009,arora:asfcs:2012,gallager:topiccorex:2017,florian:wims:2020}. 


We demonstrate the practical application of our approach by empirically assessing the topical alignment of the \rust official documentation. \rust is a programming language that is steadily gaining the attention of software engineering practitioners, being voted as the ``most loved'' language by developers for over six consecutive years~\cite{stackoverflow:devsurvey:2020}. Nonetheless, when participants of the annual Rust Survey~\cite{rust:rustsurvey:2020} are asked about what the \rust community can do to improve \rust's adoption, respondents frequently answer that training and documentation are two essential efforts. Therefore, beyond demonstrating the application of our approach, an empirical study of the topical alignment of \rust documentation provides practical recommendations to an important community of documenters.


For our case study, we encode the \rust domain knowledge into 47 different \emph{knowledge units} (KUs) that are induced from a manual categorization of the language's official documentation~\cite{rust:bookshelf:2021}. A KU represents a set of related concepts that a skilled \rust developer must know. Each KU is associated with one topic that our semi-supervised topic model derives. Each KU is also associated with a set of anchor words (e.g., the ``primitive type'' KU is associated with anchor words such as ``tuple'' and ``array'') that are given as input to the semi-supervised topic models. We then build a \emph{concrete model} from the data of Question \& Answer (Q\&A) websites~\cite{treude:icse-nier:2011,wang:sac:2013,barua:emse:2014,han:emse:2020}, which are popular venues where developers actively seek for information about programming~\cite{chen:google_so_search_compsac:2016,venkatesh:web_apis_usage:2016}. In addition, we build a \emph{documentation model} from the data of the official \rust documentation.


For both the concrete and documentation models, we analyze the \emph{prevalence of each KU} (i.e., the number of documents in which a KU occurs). Based on the prevalences, we identify the \emph{convergent KUs} (i.e., the KUs that are highly prevalent in both the concrete and documentation models), the \emph{divergent KUs} (i.e., the KUs that are highly prevalent in the concrete model but not in the documentation model), and the \emph{absent KUs} (i.e., KUs with a low prevalence in both models). We also analyze the \emph{awareness of each KU} (i.e., the number of answers that each KU receives in the Q\&A posts that are associated with the KU). Based on these analyses, we answer three research questions (RQs):

\smallskip \noindent \textbf{RQ1) \rqone}

\noindent \textit{In general, there is a strong agreement between the concrete information needs of developers and the contents of the official \rust documentation. Still, specific KUs that are associated with many Q\&A posts have a low coverage by the \rust documentation. We also observed KUs that are seldomly discussed in Q\&A websites and also have a low coverage by the official documentation.}

\smallskip \noindent \textbf{RQ2) \rqtwo}

\noindent \textit{We identified ten absent, six divergent, and ten convergent KUs in terms of the frequency of occurrence, half of the absent KUs categorized as a KU for programming niches (e.g., network, game, and database development). In addition, one-third of the KUs that are categorized as language features (e.g., structs, patterns and matchings, and foreign function interface) are divergent.}

\smallskip \noindent \textbf{RQ3) \rqthree} 

\noindent \textit{In general, absent KUs have higher ranks of attention (i.e., number of answers in Q\&A posts) and lower ranks of agreement (i.e., the odds of having an accepted answer), suggesting KUs that should be prioritized by documenters (e.g., logging and command line arguments parsing). Conversely, convergent KUs have lower ranks of attention and higher ranks of agreement.}

\smallskip

The main contributions of our paper are:

\begin{itemize}
    \item \textit{Methodological}: We propose an approach to assess the topical alignment of the documentation of a programming language using semi-supervised topic models of the posts of Q\&A websites and a language's documentation. To foster further research in the area of topical alignment of programming languages, we also provide supplementary material that contains our data and trained model.
    \item \textit{Technical}: We perform an empirical study about the topical alignment of the \rust documentation, followed by a discussion of the implications of our findings for documenters of programming languages. In our empirical study, we develop a set of metrics and visualizations to assess the topical alignment of the documentation of a programming language. Our empirical study also identifies topics that lack coverage from the \rust documentation and which of such topics should be prioritized by documenters.
    \item \textit{Conceptual}: We modelled the domain knowledge of \rust through a careful analysis of the official documentation that is maintained by the community. Our domain knowledge is validated by a specialist with years of contribution to the analyzed material (as one of the core contributors of \rust documentation). Our domain knowledge is encoded as anchor words and it can be used by a variety of semi-supervised topic modelling techniques that adopt this knowledge representation format~\cite{andrzejewski:icml:2009,arora:asfcs:2012,gallager:topiccorex:2017,florian:wims:2020}.
\end{itemize}

\noindent \textbf{Paper organization:} Section~\ref{sec:motivation} motivates our study. Section~\ref{sec:background} provides background material. Section~\ref{sec:study-design} presents our data collection procedure. Section~\ref{sec:results}, describes our results. Section~\ref{sec:discussion} discusses the implications of our findings for documenters. Section~\ref{sec:threats-to-validity} reflects on the threats to the validity of our study. Finally Section~\ref{sec:conclusions} presents our conclusions.

\section{Motivation}
\label{sec:motivation}

It is common for communities in charge of maintaining programming languages to release surveys to understand the users' opinions about diverse aspects of the language. By analyzing the latest surveys from five of the ``most loved'' languages\footnote{We did not find any survey for the \textsf{TypeScript} language, featured second on the rank of ``most loved'' languages.}~\cite{stackoverflow:devsurvey:2020} (in order, \rust, \textsf{Python}, \textsf{Kotlin}, \textsf{Go}, and \textsf{Julia}), we found that users mention the importance of documentation in all of these surveys. In the \rust survey, participants ($n = 8,323$) were asked about what they think could be done to improve the adoption of \rust, and the largest category of answers was ``documentation and training''~\cite{rust:rustsurvey:2020}. In the \textsf{Python} survey, 14\% of the respondents ($n > 28,000$) pointed ``clear documentation'' as the answer to an open question about their three favourite features of the language~\cite{python:survey:2020}. The \textsf{Kotlin} survey ($n = 1,163$) showed that almost 25\% of the respondents use the language's official documentation to set up a project. One of the conclusions drawn by the \textsf{Kotlin} community based on their survey is that ``\textit{to increase the level of satisfaction with \textsf{Kotlin Multiplatform}, we need to continue to improve and maintain the documentation}''~\cite{kotlin:survey:2021}. Participants of the \textsf{Go} survey ($n = 9,648$) were asked about what could be done to make the \textsf{Go} community more welcoming, and 21\% of the respondents of this question ($n = 275$) pointed to ``improvement of learning resources and documentation''~\cite{go:documentation_survey:2020}. In addition, other 62\% of the respondents ($n = 2,476$) pointed that they struggle to ``find enough information to fully implement a feature of my application'' in the official \textsf{Go} documentation. Similarly, in the \textsf{Julia} survey ($n=2,565$), 26\% of the respondents said that ``insufficient documentation'' is one of the biggest non-technical problems with the language~\cite{julia:survey:2020}.

The survey responses show that documentation quality is paramount in promoting the adoption and the usage of a programming language. Moreover, communities devoted to the development of a programming language are, in general, willing to improve documentation. Also, the official documentation occasionally lacks information, compelling developers to find alternative forms to obtain information. Therefore, documenters of programming languages will benefit from new approaches to understand the state of the maintained documentation, especially if these approaches are less costly than conducting surveys. As the development of the official documentation of a programming language typically depends on a centralized team and follows a push-based distribution model (i.e., documenters are responsible for the definition of the available information), each update to the documentation contents needs to be carefully considered. In particular, to maximize the utility of the documented information, documenters want to assess the alignment between the maintained documents and the information needs of developers. This requirement is exemplified in the surveys mentioned above. For example, one of the questions in the \textsf{Go} survey asks ``how helpful is official \textsf{Go} documentation for achieving your programming goals''~\cite{go:documentation_survey:2020}? Also, one of the immediate actions stated by the \textsf{Kotlin} community in response to the results of their documentation-related questions includes ``\textit{focus (...) on common user scenarios that are not well covered in documentation}''~\cite{kotlin:survey:2021}. 

In this paper, we describe a machine learning-based approach that documenters can adopt to assess the alignment between the concrete information needs of developers (represented by related posts in Q\&A websites) and a programming language's documentation (represented by the officially maintained documentation). The information needs from developers are represented by the topics derived from Q\&A websites, as this type of documentation follows a pull-based distribution model (i.e., information seekers are responsible for the definition of the available information), and developers post legitimate questions to obtain voluntary answers from other developers. Our approach, described in detail in Section~\ref{sec:study-design:subsec:measuring-concept-align}, uses semi-supervised topic modelling that leverages domain knowledge to cluster documents into topics (each topic associated with one KU from \rust). We first build a concrete model using data from \ rust-related Q\&A posts, then we use this model to derive the topics of the official \rust documentation. A set of KU metrics, described in detail in Section~\ref{sec:study-design:subsec:calculate-ku-metrics}, are extracted from the built models and compared against each other, providing documenters with information about the similarities between the concrete and the documentation models in terms of these KUs.
\section{Background and related works}
\label{sec:background}

This section discusses the background material of our study, as well as related works. Section~\ref{sec:background:subsec:documentation-pl} presents prior studies about programming language documentation and software documentation in general. Section~\ref{sec:background:subsec:qa-websites} presents prior works that extracted developer's information needs from Q\&A websites. Section~\ref{sec:background:subsec:topic-modelling} presents a semi-supervised topic modelling technique called Anchored Correlation Explanation (CorEx), and discusses the advantages of applying this specific technique over unsupervised techniques to mine textual data.

\subsection{Programming languages documentation for software engineering}
\label{sec:background:subsec:documentation-pl}

In the context of software engineering, programming language documentation refers to the technical information that communicates to developers concepts about the \emph{usage} of a programming language (i.e., high-level concepts), in contrast with its \emph{implementation} (i.e., formal definitions)~\cite{shaw:jovial_cacm:1963}. Examples of such documents include manuals, training material, and API references. The maintenance of programming language documentation is deemed as a fundamental practice since the initial days of software engineering. In 1963, refering to the documentation of the \textsf{IPL-V} programming language (released a few years earlier), Allen~\cite{allen:doc_ipl_v:cacm} wrote that ``\textit{the main documentation of IPL-V can be considered complete, official and almost permanently fixed}'' and that the concern of future updates to the language documentation was addressed: ``\textit{(...) documentation is kept on tape with procedures for updating and modifying the documentation}''. Yngve et al.~\cite{Yngve:communacm_pl_doc:1963} described some of the pioneering research in programming languages documentation, recognizing the importance and discussing the challenges of maintaining this type of documentation. Interestingly, more recent research recognizes (general) documentation maintenance as an often overlooked software engineering practice~\cite{lethbridge:how_se_use_documentatio:2003,mattsson:2005}, pointing to the cost of maintaining documentation as one of the drivers of this scenario. 

At the intersection of programming languages documentation and software engineering automation, researchers proposed different approaches to automatically generate documentation, often by analyzing the structure of source code. Such automated approaches aim to supporting different development tasks (such as exception handling~\cite{buse:auto_doc_infer_exceptions:2008}) and different niches (such as scientific programming~\cite{moser:doc_generator_sci_eng:2015}). Automated approaches are also used to generate documentation to different features of the language, such as its API~\cite{mcburney:aut_doc_gen_via_source_code:2014, zhai:aut_model_gen_doc_java_api:2016,souza:ist:2019}, components~\cite{moreno:auto_gen_nlp_java_class_summary:2013}, and specific syntatic expressions~\cite{anwar:auto_lambda_java_doc:2019}. One of the main challenges faced by automated documentation tools is to capture the information needs of developers~\cite{ko:how-devs_seek_info:2006, ko:info_needs_collocated_teams:2007, treude:extracting_dev_tasks:2015}. For example, Robillard et al.~\cite{robillard:on_demand_doc:2017} discussed that research in automated documentation should produce results that better support the information needs of developers. Treude et al.~\cite{treude:extracting_dev_tasks:2015} proposed an automated tool to extract tasks from documentation to help bridging the gap between the documentation structure and information needs of developers. Also, the increasing popularity of Q\&A websites as a source of documentation has been explored by researchers to capture the information needs of developers~\cite{chen:google_so_search_compsac:2016,yuhao:how_devs_use_so_code:2019}.

Nonetheless, most of the existing approaches to automate programming language documentation do not generate information that directly supports the decision-making process of designers, producers, and curators of documentation (i.e., documenters). Instead, most of the existing automated approaches focus on supporting program comprehension by developers (i.e., the users of the documentation)~\cite{uri:reading_doc_for_program_comprehension:2009}. Although the automatic generation of documentation~\cite{xing:deep_code_comment_generation:2018} and the identification of information needs of developers can be leveraged to support documentation activities, documenters have different requirements than developers that passively consume the documentation. For example, due to limited resources, documenters need to prioritize documentation effort~\cite{mcBurney:prioritizing_documentation:2018}. As a consequence, automated techniques to support documenters' decision-making are necessary, particularly those techniques that are focused on external documenters that produce material about the \emph{usage and features of a programming language}, without any particular association with a specific software system. Such documenters, as described in Section~\ref{sec:motivation}, are typically organized as dedicated teams around a specific programming language. 

\subsection{Topic Modelling Q\&A websites}
\label{sec:background:subsec:qa-websites}

Q\&A websites (e.g., \stkovflw) are popular venues where developers discuss technical aspects of programming. These websites allow users to post \emph{questions} about a topic and to obtain \emph{answers} from other users. A \emph{post} in a Q\&A website combines the question and the associated answers. Such information has been leveraged by software engineering researchers to understand what developers discuss about a variety of technologies. Information from posts are used to investigate the topic trends and challenges in areas such as mobile development~\cite{rosen:emse:2016}, security~\cite{yang:jcst:2016}, blockchain~\cite{wan:tse:2019}, machine learning~\cite{bangash:msr:2019}, big data~\cite{bagherzadeh:fse:2019}, deep learning frameworks~\cite{han:emse:2020,chen:fse:2020}, configuration as code~\cite{rahman:rcose:2018}, concurrency~\cite{ahmed:esem:2019}, internet of things~\cite{aly:iot:2021}, chatbots~\cite{abdellatif:msr:2020}, and new programming languages~\cite{chakraborty:ist:2021}.

In prior works, the discussion topics in Q\&A websites are learned via unsupervised topic models trained over the textual data of the posts~\cite{barua:emse:2014}. Topic models learn two hidden components from the data of Q\&A websites. The first component is the set of topics, with each topic being described as a probability distribution of word types. The second component is the set of documents, with each document being described as a probability distribution of topics (each document corresponds to a unique post). By associating topics with documents (posts), researchers leverage the metadata of the posts to calculate metrics that denote the popularity and the difficulty of the topics~\cite{ahmed:esem:2019,alshangiti:esem:2019,bagherzadeh:fse:2019,abdellatif:msr:2020} (e.g., the popularity of a topic is measured as the number of times a topic is associated with a post).

The usage of unsupervised topic modelling techniques incurs the cost of topic labelling. After using an unsupervised technique to extract the set of topics from the data, researchers must manually inspect the words with the largest probability to label a topic. However, it is not trivial to systematically assign labels to topics, and this task typically involves manually analyzing the top words in each topic, measuring the inter-rater agreement between researchers~\cite{chakraborty:ist:2021}, merging related topics~\cite{bagherzadeh:fse:2019}, and discarding meaningless topics~\cite{bangash:msr:2019}. Existing efforts to systematically assign labels to topics involve the derivation of reference architectures and the posterior mapping of topics to layers of the architecture~\cite{wan:tse:2019,han:emse:2020}. Still, topics can be difficult to be interpreted, and some are especially hard to make sense of due to ambiguities. For example, Hindle et al.~\cite{hindle:icsm:2012} shows that domain experts are more likely to produce accurate topic labels than non-experts. Moreover, topic models are susceptible to over-representation of frequent (i.e., more general) topics -- an effect called ``rich topics get richer''~\cite{fang:chinacommunications:2014}. As a result, unsupervised techniques hardly represent all the topics that agree with the researchers' knowledge about the studied domain, especially when topics of interest are underrepresented in a corpus. To address this challenge, one can adopt a semi-supervised topic model that uses encoded domain knowledge to bias the derivation of topics towards topics of interest.

\subsection{Topic modelling with domain knowledge}
\label{sec:background:subsec:topic-modelling}


Anchored CorEx~\cite{gallager:topiccorex:2017} is a semi-supervised topic modelling technique that does not assume a probabilistic generative model and avoids the need of adjusting parameters of a prior probability distribution such as in LDA-based topic models~\cite{blei:icml:2006}. Anchored CorEx also accepts the integration of domain knowledge in the form of \emph{anchor words}, i.e., a set of keywords that are associated with the topics of interest. These anchor words act as markers that help the model to distinguish particular topics from others.\footnote{Anchor words is a related concept to prototypes in general clustering~\cite{kuncheva:cluster_prototype:1998}.} Therefore, the generated topics can be pre-labelled according to the assigned meaning to their associated set of anchor words.

\subsubsection{Anchored Correlation Explanation}

The CorEx approach to topic modelling finds the set of topics that better ``explain'' the dependence (a.k.a. correlation) between each topic (a group of words) and the documents in the data~\cite{versteeg:nips:2014}. Consequently, the learned parameters by CorEx are interpreted as the ``association'' between the topic and a document. Estimating dependence is a powerful characteristic of CorEx to represent topics without having to adjust any distribution parameter. The generated topic models by CorEx assign an individual weight for each topic in a document (i.e., the sum of the topic probabilities vary from 0 to $m$, where $m$ is the number of topics). In the CoreEx theory, the \emph{total correlation} is expressed as $TC(X_G; Y) = \sum_{i \in G}{I(X_i: Y) - I(X_G: Y)}$, where $I$ is the mutual information of two random variables, $X_i$ is a word type, $X_G$ is a specifc subset of all word types $G$, and $Y$ represents a topic. CorEx finds the configuration of topics that are the ``most informative'' of the words and documents by maximizing $TC(X_j; Y_1, \dots, Y_m)$, where $Y_1, \dots, Y_m$ are topics and $X_j$ the corresponding set of word types of each topic, for $j=1, ..., m$. Each topic contributes differently to the total correlation of the model. In turn, each document of the corpus contributes differently to the total correlation of a topic.

An extension to CorEx is Anchored CorEx~\cite{gallager:topiccorex:2017}, which accepts the incorporation of domain knowledge into the topic models. A set of anchor words~\cite{arora:asfcs:2012} representing topics of interest is input to the model, with different anchoring strategies producing different results in terms of topic derivation. When anchoring for topic separability, the best strategy is to assign a set of (potentially exclusive) anchor words to each topic. During training, the association between anchor words and each topic is emphasized by a parameter $\beta$ that controls the anchor strength. The ability to incorporate anchor words to nudge the topic's derivation is a valuable feature of CorEx, as it produces topics that capture a variety of related concepts to \rust and that are of interest to documenters. 

In our study of the conceptual alignment of \rust documentation, the ability to cluster posts into topics is one of the most critical quality attributes of a topic model, since the studied metrics (described in Section~\ref{sec:study-design:subsec:calculate-ku-metrics}) are based on the probability of a topic in a document. Regarding the ability to cluster documents into topics (by associating a document to the topic with the highest probability), Anchored CorEx is shown to be better or comparable to other semi-supervised topic models in two benchmark datasets (Disaster Refief Articles and 20 News Group)~\cite{gallager:topiccorex:2017}, with respect both to the intracluster quality (homogeneity) and the inter clusters quality (adjusted mutual information).

\section{Approach}
\label{sec:study-design}

\begin{figure}
    \centering
    \includegraphics[scale=0.45]{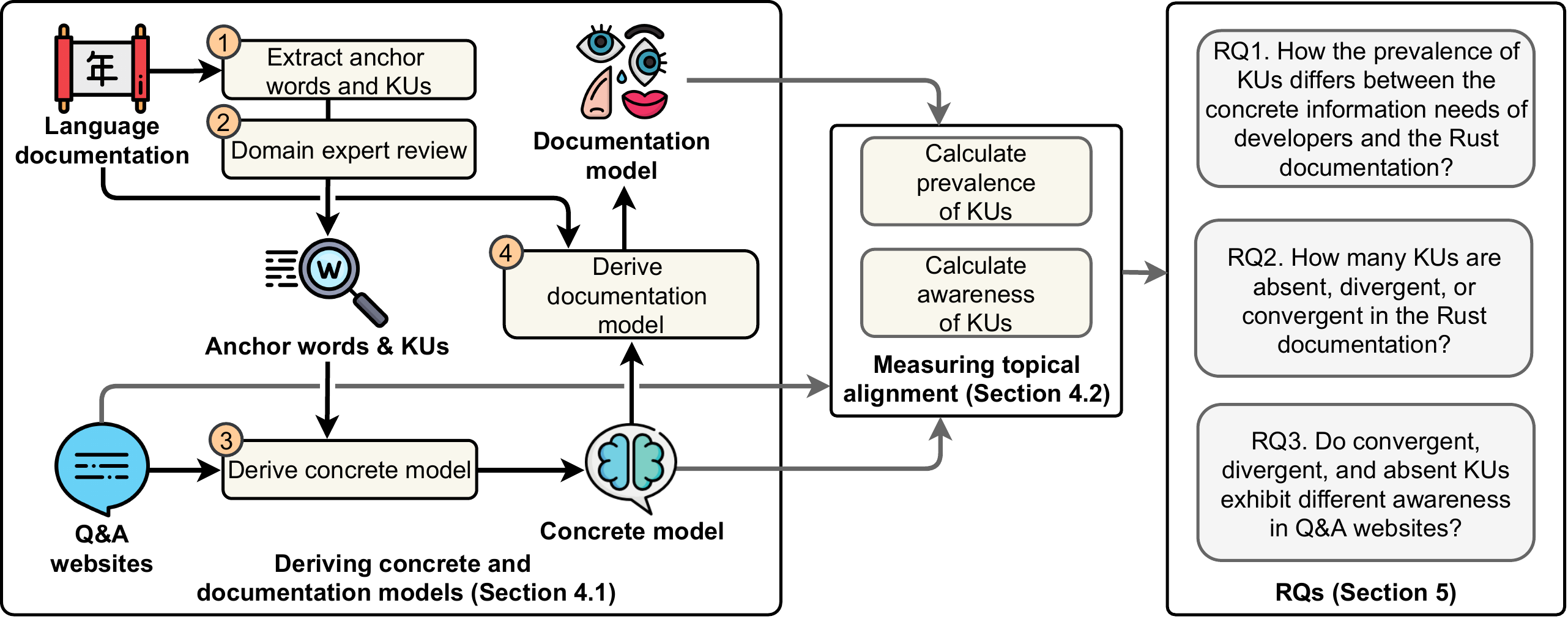}
    \caption{The overview of our approach to studying the topical alignment of \rust documentation.}
    \label{fig:approach-overview}
\end{figure}

This section describes our approach to assess the topical alignment of the documentation of programming languages (Section~\ref{sec:study-design:subsec:measuring-concept-align}) and to calculate metrics for our empirical study of the topical alignment of \rust documentation (Section~\ref{sec:study-design:subsec:calculate-ku-metrics}). Figure~\ref{fig:approach-overview} depicts the overview of our approach.

\subsection{Deriving concrete and documentation models}
\label{sec:study-design:subsec:measuring-concept-align}

We adopt three main steps to derive the concrete and the documention models. We first extract anchor words and KUs from the language documentation (Section~\ref{sec:study-design:subsec:measuring-concept-align:subsubsec:anchor-words-ku}) and review the extraction results with a domain expert. In the next step, we derive a semi-supervised topic model from Q\&A websites using the extracted anchor words to guide the topics derivation (Section~\ref{sec:study-design:subsec:measuring-concept-align:subsubsec:train-concrete-model}). Afterwards, we derive the topics of the official language documentation by leveraging the topics of the concrete model that was built in the previous step (Section~\ref{sec:study-design:subsec:measuring-concept-align:subsubsec:predict-abstract-model}).

\subsubsection{Extract anchor words and KUs:}
\label{sec:study-design:subsec:measuring-concept-align:subsubsec:anchor-words-ku}

\hfill

\smallskip A KU is defined as a group of thematically related concepts within a general knowledge domain (e.g., a programming language). In the knowledge domain of programming languages, each KU encapsulates a set of related concepts that need to be understood by developers. For example, developers need to understand how to operate different \emph{primitive data types} (e.g., scalars, arrays, and tuples) to properly use the language. Hence, primitive data types can be identified as a KU of a programming language. The documentation of a programming language describes KUs such that developers can understand and operationalize the different concepts of the language.

To identify KUs for our case study on \rust, we thoroughly read the official documentation of the language and categorize its contents into KUs. More specifically, we analyze the contents of 8 documents~\cite{rust:bookshelf:2021} that cover the major relevant concepts of the language and are officially distributed with the standard \rust installation (in particular, we analyzed the documentation that is distrbuted with \rust version 1.49.0), namely The Book~\cite{rust:book:2021}, The Reference~\cite{rust:reference:2021}, The Cargo Book~\cite{rust:cargobook:2021}, The \textsf{rustc} Book~\cite{rust:rustcbook:2021}, The \textsf{rustdoc} Book~\cite{rust:rustdocbook:2021}, Rust by Example~\cite{rust:rustbyexample:2021}, Nomicon~\cite{rust:rustonomicon:2021}, and the Edition Guide~\cite{rust:editionguide:2021}. During the categorization of the \rust documentation into KUs, we also extract a set of associated keywords with each KU. Such keywords are used as the anchor words of our semi-supervised topic models.

\subsubsection{Domain expert review:}
\label{sec:study-design:subsec:measuring-concept-align:subsubsec:domain-expert-review}

\hfill

\smallskip We validate the extracted KUs and anchor words from the \rust documentation with a domain expert that actively participates in the \rust documentation community\footnote{\url{https://prev.rust-lang.org/en-US/community.html}} for several years. The domain expert suggested the creation of three additional KUs (and their respective anchor words), as well as the splitting of another KU and additional anchor words for some of the KUs that were originally derived. At the end of this process, we ended up with 47 KUs and their associated anchor words (the associated anchor words with each KU are described in Table~\ref{tab:kus-anchor-words} of Appendix~\ref{apx:kus-description}). The KUs are grouped into four categories and shown in Table~\ref{tab:ku-category-description}.





\begin{table}
    \centering
    \caption{Derived KUs from the official \rust documentation and their respective categories.}
    \label{tab:ku-category-description}
    \footnotesize
    \begin{tabular}{p{0.16\freewidth}p{0.25\freewidth}p{0.55\freewidth}}
    \toprule
    \textbf{Category} & \textbf{Description} & \textbf{Knowledge Units} \\* \cmidrule{1-3} 
    
    Data types & Data structures that are natively provided by the \rust language. & Advanced types, Collections, Enumeration, Generics, Primitive types, Smart pointers, and Structs \\
    
    Development tooling & Tools to support the development of \rust applications.  &  Build, Compilation, Debugging, Dependency management, Documentation, IDE, Installation and setup, Modularization, Testing, and Linters \\
    
    Language features & Functionalities and attributes that are natively provided the the \rust language. & Aliasing, Advanced Functions and Closures, Casting, Control flow, Distribution channels, Exception handling, File I/O, Foreign Functional Interface, Functional language features, Lifetime, Macros, Mutability, Object orientation, Operators and symbols, Ownership, Patterns and matchings, Traits, and Unsafe Rust \\

    Programming niche & Specialized set of features that suits particular applications of \rust. & Audio, Async, Command line arguments parsing, Command line interface, Concurrency, Database, Embedded development, Formatted print, Game development, Logging, Network development, and Web development \\
    \bottomrule
    \end{tabular}
\end{table}

\subsubsection{Derive concrete model:}\label{sec:study-design:subsec:measuring-concept-align:subsubsec:train-concrete-model}

\hfill

\begin{figure}
    \centering
    \includegraphics[scale=0.45]{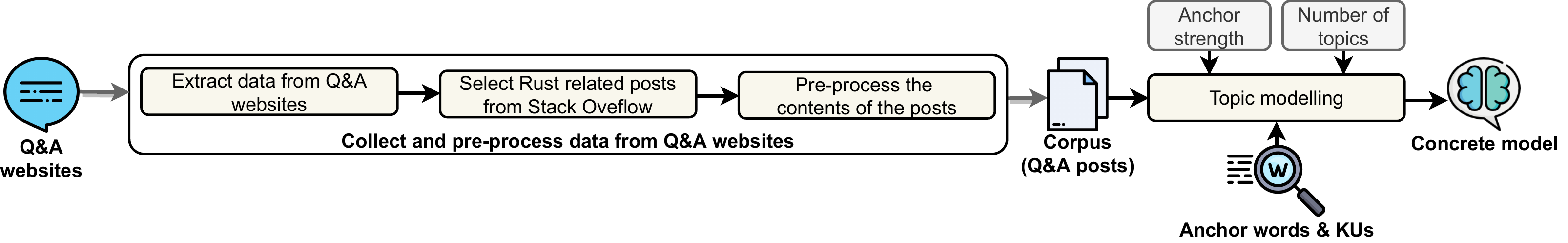}
    \caption{The performed steps to derive the concrete model.}
    \label{fig:train-concrete-model}
\end{figure}

\smallskip The concrete model of a programming language documentation represents the legitimate information needs of developers, as expressed by the voluntary discussions in Q\&A websites. Figure~\ref{fig:train-concrete-model} shows the performed steps to derive the concrete model for our \rust case study. We initially \emph{collect and proccess data} from Q\&A websites to generate our corpus. We then use Anchored CorEx~\cite{gallager:topiccorex:2017} (see Section~\ref{sec:background:subsec:topic-modelling}) to derive topics from the corpus, setting the anchor strength and number of topics parameters. We use the extracted anchor words of Section~\ref{sec:study-design:subsec:measuring-concept-align:subsubsec:anchor-words-ku} as prototypes~\cite{haghighi:prototype_driven:2006,florian:wims:2020} to each of the derived topics. In the following, we describe how we collect data from Q\&A websites and how we derive our topic model.

\smallskip \noindent \textbf{Collect and pre-process data from Q\&A websites.} Our data collection process encompasses three steps: \emph{extract data from Q\&A websites}, in which we obtain data from \stkovflw~-- a popular Q\&A website for general programing related discussions~-- and \rustforum~-- a specific Q\&A website maintained by the \rust community, \emph{select \rust related posts of \stkovflw}, in which we select \rust related posts from \stkovflw, and \emph{pre-process the contents of the posts}, in which prepare our data for our topic modelling.

\smallskip \noindent \textit{Extract data from Q\&A websites:} We extract metadata about the posts of two Q\&A websites: \stkovflw and \rustforum. Data from \stkovflw is publicly available through the \textsf{Stack Exchange Data Explorer} \cite{stackexchange:dataexplorer:2021} tool, from which we collect the \texttt{Posts.xml} file containing metadata about the posts (the file was downloaded on January 11th, 2021). Each post entry consists of either a question or one of its associated answers. We match the \texttt{Id} field of a question (i.e., an entry whose \texttt{PostIdType} field is equal to 1) with the \texttt{ParentId} field of an answer (i.e., an entry with \texttt{PostIdType} equal to 2) to link the question and the associated answers that belong to the same post. Table~\ref{tab:collected-metadata} describes the collected metadata of each post in \stkovflw. 

The \rustforum website is built upon the open-source \textsf{Discourse}~\cite{discourse:website:2021} platform, which allows data to be collected using an API~\cite{discourse:apidocumentation:2021}. We start by collecting all the questions\footnote{We access the \url{https://users.rust-lang.org/c/help/5.json?page={page_number}} API endpoint, where \texttt{page\_number} is a placeholder to paginate the results.} under the ``help'' category (i.e., the category of questions that are meant to request help from users of the forum). We select questions of the ``help'' category because this is the category under which developers discuss technical issues about the usage of \rust to solve programming problems and makes up more than 65\% of all posts (the other two larger categories of posts include ``uncategorized'' and ``announcements'' that, together with ``help'', make up more than 92\% of all posts). After obtaining metadata of all questions, we collect the metadata of all answers\footnote{We access the \url{https://users.rust-lang.org/t/-/{topic_id}.json?page={page_number}} API endpoint, where \texttt{topic\_id} is a placeholder for the question \texttt{id} field and \texttt{page\_number} is a placeholder to paginate the results} associated with each question. Data from \rustforum was collected on February 9th, 2021. Similar to the data from \stkovflw, a post combines the original question and its associated answers. We use the \texttt{id} field of a question and the \texttt{topic\_id} field of an answer to link the question and the associated answers of the same post. In total, we collected $14,520$ posts from \rustforum along with $81,225$ answers. Table~\ref{tab:collected-metadata} describes the collected metadata of each post in the \rustforum and on \stkovflw.

\begin{table}
    \centering
    \caption{The posts metadata collected from \stkovflw and \rustforum.}
    \label{tab:collected-metadata}
    \footnotesize
    \begin{tabular}{p{0.16\freewidth}p{0.2\freewidth}p{0.3\freewidth}p{0.3\freewidth}}
    \toprule
    \textbf{Metadata}        & \textbf{Description}                                                                                   & \textbf{Stack Overflow}                                                                                                                                                                       & \textbf{Rust Forum}                                                                                                                                              \\ \midrule
    Post contents            & The textual information of the question and answers. & The \texttt{Body} field of the question and the associated answers. & The \texttt{cooked} field of the question and the associated answers. \\
    
    Post title               & The associated title with the post. & The \texttt{Title} field of the question.                                                                                                & The \texttt{title} field of the question.                                                                   \\
    
    
    Accepted answer date     & The date of the accepted answer of the post.                & The \texttt{CreationDate} field of the answer whose \texttt{Id} field matches the \texttt{AcceptedAnswerId} field of the question. & The \texttt{created\_at} field of the answer whose \texttt{accepted\_answer} field is true.              \\
    
    Number of answers        & The number of associated answers with the post. & The number of \texttt{ParentId} fields of the answers that match the \texttt{Id} field of the question. & The number of \texttt{topic\_id} fields of the answers that match the \texttt{id} field of the question. \\
    
    
    Tags & The user assigned tags to the post. & The \texttt{Tags} field. & -- \\ \bottomrule
    \end{tabular}%
    \end{table}

\smallskip \noindent \textit{Select \rust related posts of \stkovflw:} In \stkovflw, developers can discuss the solution of problems related with any programming language. Since we collect all posts from \stkovflw, we need to select posts that are related to \rust. To this end, we leverage the user-assigned tags to identify all posts of interest. In particular, we search for all tags in \stkovflw that have ``rust'' as a prefix, ending up with a total of 12 tags (including the ``rust'' tag). In addition, we derive a set of keywords from the language's documentation \cite{rust:bookshelf:2021} that \stkovflw users could potentially use to tag a \rust-related post. We identified ambiguous keywords that could be used to tag posts related with other programming languages (e.g., ``generics'', ``closure'', and ``enum''). For sanity checking, we manually analyze all tags of the posts that are tagged with any of the derived keywords, then we remove keywords that are associated with unrelated posts. For instance, many posts with the tag ``generics'' also have the tag ``java'' and, for this reason, we remove ``generic'' from our initial tag set. Table~\ref{tab:initial-set-rust-tags} in Appendix~\ref{apx:initial-tag-set} shows our initial set of \rust related tags.

After determining the initial set of keywords, we adopt a similar approach to prior works \cite{linares-vasquez:msr:2013,rosen:emse:2016,wan:tse:2019} to expand the initial set of tags and to remove irrelevant tags. More specifically, to expand the initial set of tags, we first search for all posts that contain any of our initial tags of Table~\ref{tab:initial-set-rust-tags}, from which we identify an additional 2,887 tags that are associated with those posts. To remove any irrelevant tags out of the 2,887 tags, we perform a filtering process based on two thresholds, namely \emph{tag exclusivity threshold} ($\text{TET}$) and \emph{tag significance threshold} ($\text{TST}$):

\begin{align*}
    \text{TET}_{tag} &= \frac{\text{Number of Rust posts}_{tag}}{\text{Number of total posts}_{tag}} > 50\% \\
    \text{TST}_{tag} &= \frac{\text{Number of Rust posts}_{tag}}{\text{Number of Rust posts with the most popular tag}} > 1\%
\end{align*}

\noindent where:

\begin{conditions*}
\text{Number of Rust posts}_{tag} & number of posts tagged with both $tag$ and any of the initial \rust related tags. \\
\text{Number of total posts}_{tag} & number of posts tagged with $tag$. \\
\text{The most popular tag} & ``rust'', which tags $19,335$ posts tagged with any of the initial \rust related tags.
\end{conditions*}

The rationale for filtering out tags with $\text{TET}_{tag} <= 50\%$ is to remove tags that are not exclusive of \rust related posts. The rationale for filtering out tags with $\text{TST}_{tag} <= 1\%$ is that tags with high $\text{TET}$ can be associated with a very low number of \rust related posts. In these cases, $\text{TST}$ will indicate a low confidence level in the tag's significance to denote a post of interest. We manually inspect the tags that survive to our threshold criteria and select the tags that are exclusively related to \rust. In addition to the tags shown in Table~\ref{tab:initial-set-rust-tags}, three other tags are added to our tag set during this manual inspection process: ``lifetime'', ``serde'', and ``borrowing''. By selecting the posts with any of these tags, we ended up with $19,603$ posts along with $22,715$ answers. Formally, we denote the set of selected posts in each Q\&A website $q$ as $P_q$.

\smallskip \noindent \textit{Pre-process the contents of the posts:} To prepare our data for our topic modelling, we concatenate the post title and the post contents (see Table~\ref{tab:collected-metadata}) of each Q\&A website. We then combine the posts of both Q\&A websites in a unique set of posts. Finally, we perform six pre-processing steps:

\begin{enumerate}
    \item Remove code snippets within \texttt{<code></code>}.
    \item Remove HTML tags.
    \item Remove punctuation and non-alphabetic characters.
    \item Replace compound anchor words with their underscored format.
    \item Lemmatize.
    \item Remove stop words.
\end{enumerate}

To remove HTML tags, we use the \textsf{BeautifulSoup}\footnote{\url{https://pypi.org/project/beautifulsoup4/}} library. We replace any occurrence of a compound anchor word in a post by its underscored format. For example, whenever we identify the occurrence of the ``primitive type'' anchor word in a post, we replace such an occurrence with ``primitive\_type''. The provided anchor words to the topic modelling technique are also replaced by their underscored format. The rationale for this preprocessing step is to ensure that compound anchor words can be properly captured during the derivation of topics. For lemmatization and stop word removal, we used the \textsf{NLTK}\footnote{\url{https://www.nltk.org/}} library. After performing the six aforementioned pre-process steps, we tokenize the contents of the posts and obtain our corpus for the Q\&A websites (see Figure~\ref{fig:train-concrete-model}).

\smallskip \noindent \textbf{Topic modelling.} We use the Anchored CorEx implementation~\cite{corex:implementation:2021} to automatically derive topics from the contents of the \rust related posts and to assign a set of topics to each post. Two parameters are given to the topic modelling technique: the number of topics and the anchor strength (i.e., the probability weight assigned to the anchor words in relation to the other words in a topic).

\begin{figure}
    \includegraphics[scale=0.35]{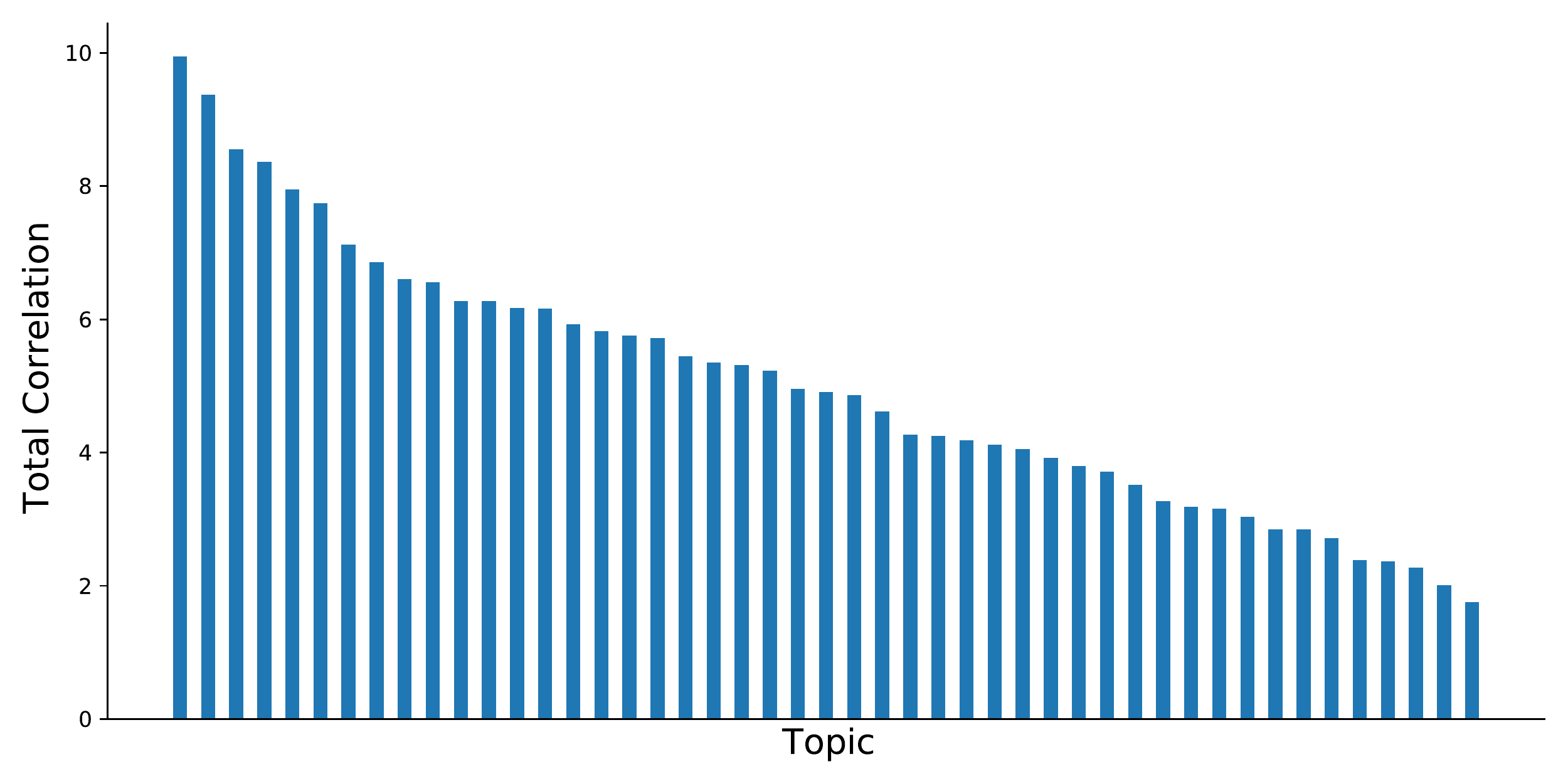}
    \caption{The distibution of the topic TC of the model with 47 anchored topics.}
    \label{fig:topic-total-correlation}
\end{figure}

\smallskip \noindent \textit{Number of topics:} Ideally, our topic models should have the same number of topics as the number of derived KUs, since this characteristic ensures a one-on-one mapping of anchor words and topics, which facilitates our analyses in terms of KUs. We perform two steps to determine whether we can obtain quality topics in a model that has the same number of topics as KUs (or whether, instead, more topics would be needed). In our first step, we derive two tentative topic models, one with 47 anchored topics (one topic per KU) and another model with the same 47 anchored topics and an additional non-anchored topic. We then manually analyze the additional non-anchored topic to verify the interpretability of this topic. We find that this additional topic is composed of commonly used words by developers in Q\&A websites that cannot be directly related to a KU of \rust. Specifically, the top-5 words in this topic are think, make, thing, case, and need. In a second step, we decompose the models' overall total correlation (TC) (see Section~\ref{sec:background:subsec:topic-modelling}) into a distribution of individual topic TCs that are calculated during model training~\cite{versteeg:nips:2014}. If we observe relatively small values of topic TC in this distribution (in relation to the higher values of TC), then the number of topics of the model is sufficient to explain the correlation in the data, and additional topics will not contribute to the overall TC~\cite{gallager:topiccorex:2017}. Figure~\ref{fig:topic-total-correlation} shows that the model with 47 anchored topics presents a good compromise between the ability to explain the correlation in the data without additional topics and the requirement of one-on-one mapping of anchor words and topics. Assuming that a one-on-one mapping exists between anchor words and topics, we label a topic with the KU that is associated with the set of anchor words of that topic. For this reason, during our analysis of the topic model (Section~\ref{sec:results}), we use the terms KU and topic interchangeably. In our formal definitions, we denote the set of topics as $K$.

\smallskip \noindent \textit{Anchor strength:} To determine the anchor strength (see Section~\ref{sec:background:subsec:topic-modelling}), we tentatively build models with anchor strength parameters $\beta = \{2,5,10,15\}$. For each of the tentative models, we manually inspect the generated topics and qualitatively evaluate how well the top 20 words in the topic agree with the associated KU with that topic (via assigned anchor words). Based on this qualitative evaluation, we conclude that an anchor strength parameter $\beta = 10$ produces the best topic-word distribution among the tentative models. The 47 derived topics are shown in Table~\ref{tab:kus-anchor-words} of Appendix~\ref{apx:topics}. Deriving the concrete model results in a topic model $M_{\text{Concrete}}$ with estimate probabilities $p(y|x)$ of a certain topic $y$ to occur in a document (i.e., a Q\&A post) $x$, given the words in $x$. 

\subsubsection{Derive documentation model:}\label{sec:study-design:subsec:measuring-concept-align:subsubsec:predict-abstract-model}

\hfill

\begin{figure}
    \centering
    \includegraphics[scale=0.45]{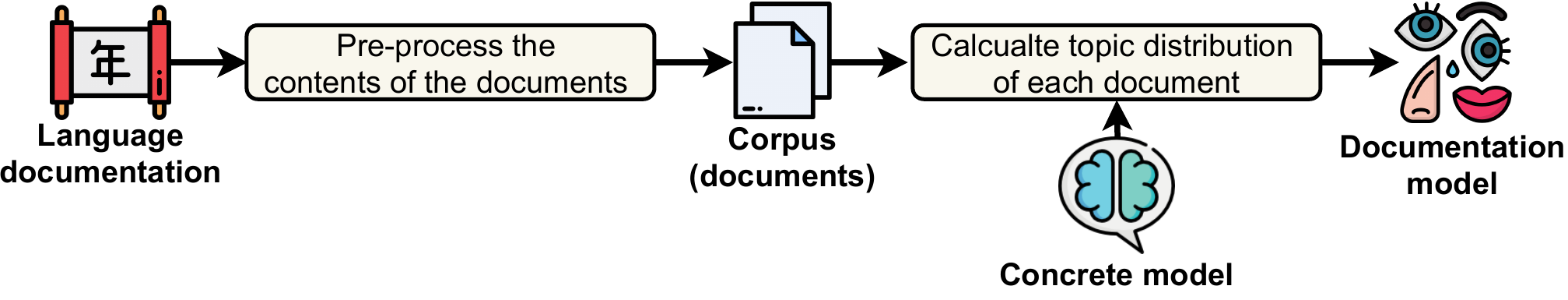}
    \caption{The performed steps to derive the documentation model.}
    \label{fig:train-documentation-model}
\end{figure}

\smallskip The derivation of the documentation model encompasses the two depicted steps of Figure~\ref{fig:train-documentation-model}. First, we \emph{pre-process the contents of each document} of the official \rust documentation, from which we obtain a corpus of the documentation. Next, we \emph{calculate the distribution of topics that occur in each document} of the corpus.

\noindent \textbf{Pre-process the contents of the documents.} The official \rust documentation consists of a series of books. In turn, each book consists of a set of chapters that cover different aspects of the language. To build the corpus of the official language documentation, we collect each chapter of the eight documents described in Section~\ref{sec:study-design:subsec:measuring-concept-align:subsubsec:anchor-words-ku}, in addition to the chapters of three reference API documents, namely The Rust Core Allocation and Collections Library~\cite{rust:allocbook:2021}, The Rust Core Library~\cite{rust:core:2021}, and the Rust Standard Library~\cite{rust:std:2021}. These eleven books represent the complete documentation distributed with the standard installation of \rust version 1.49.0. The books are distributed in a standard \texttt{html} format that is automatically generated by the \textsf{mdBook} tool\footnote{\url{https://github.com/rust-lang/mdBook}}. To pre-process the contents of each book chapter, we initially parse the respective \texttt{html} files and extract the text within any \texttt{<p></p>} tags. We then concatenate all the extracted text that is associated with the same \texttt{html} file (as each file contains the material of one book chapter). Finally, we apply the same preprocessing steps described in Section~\ref{sec:study-design:subsec:measuring-concept-align:subsubsec:train-concrete-model} and tokenize the extracted text from the \rust documentation. In total, our corpus contains $4,754$ documents, each one corresponding to a book chapter.\footnote{In total, there are $4,771$ chapters in the books of the official \rust documentation. However, $17$ of such chapters do not contain any textual information within \texttt{<p></p>} tags, leaving us with a total of $4,754$ useful chapters.} We then use the \rust documentation corpus and the concrete model (see Section~\ref{sec:study-design:subsec:measuring-concept-align:subsubsec:train-concrete-model}) to build our documentation model.

\noindent \textbf{Calculate the topic distribution of each document.} The documentation model of a programming language represents the topics that occur in each document of the language's documentation. By deriving the documentation model, our objective is to measure the topical alignment of \rust documentation by comparing the prevalence of the topics of the concrete model against the prevalence of the topics of the documentation model (further details about the calculation of topic prevalence is given in Section~\ref{sec:study-design:subsec:calculate-ku-metrics}). The comparison between topics of the concrete and the documentation model is based on two assumptions. Our first assumption is that we only compare topics that occur in both models. As we derive the concrete model using induced anchor words from the language's documentation, topics that occur in the concrete model also occur in the documentation by definition. Hence, we need to suppress topics that occur exclusively in the documentation model. The suppression of exclusive topics of the documentation model does not limit the measurement of topical alignment, as we are not interested in making claims regarding topics that occur only in the documentation. For instance, we do not suggest removing topics that are not mentioned in any Q\&A post from the documentation, as a missing topic from Q\&A websites might be due to the wide coverage of that topic by the documentation. Our second assumption is that we perform a paired comparison of topics between the Q\&A posts and the official documentation. By paired comparison, we mean comparing the same topics (word types distribution) between the concrete and the documentation models.

To satisfy our two assumptions, we leverage the $M_{\text{Concrete}}$ model to calculate the probability of each topic from $M_{\text{Concrete}}$ to occur in each document of the official documentation collection. This procedure contrasts with the derivation of a new semi-supervised topic model from data of the official language's documentation, potentially containing different topics than the topics of the concrete model. To derive the documentation model, we apply the same optimization procedure as described in Section~\ref{sec:background:subsec:topic-modelling} to assign topics to each document of the language's official documentation. More specifically, for each document of the official documentation, we calculate the configuration of topics $Y_{\text{Concrete}}$ from the concrete model that maximizes $TC(X_{\text{Documentation}};Y_{\text{Concrete}})$, where $X_{\text{Documentation}} \subseteq  X_{\text{Concrete}}$ (i.e., the word types of the documentation model is constrained to the word types of the concrete model). We use the \texttt{predict} function provided by CorEx~\cite{corex:implementation:2021} as the implementation of our optimization procedure. The output of this procedure is the probability of each topic $y$ of $Y_{\text{Concrete}}$ in each document $x$ of the official documentation. 

\subsection{Measuring topical alignment}
\label{sec:study-design:subsec:calculate-ku-metrics}

\begin{figure}
    \centering
    \includegraphics[scale=0.45]{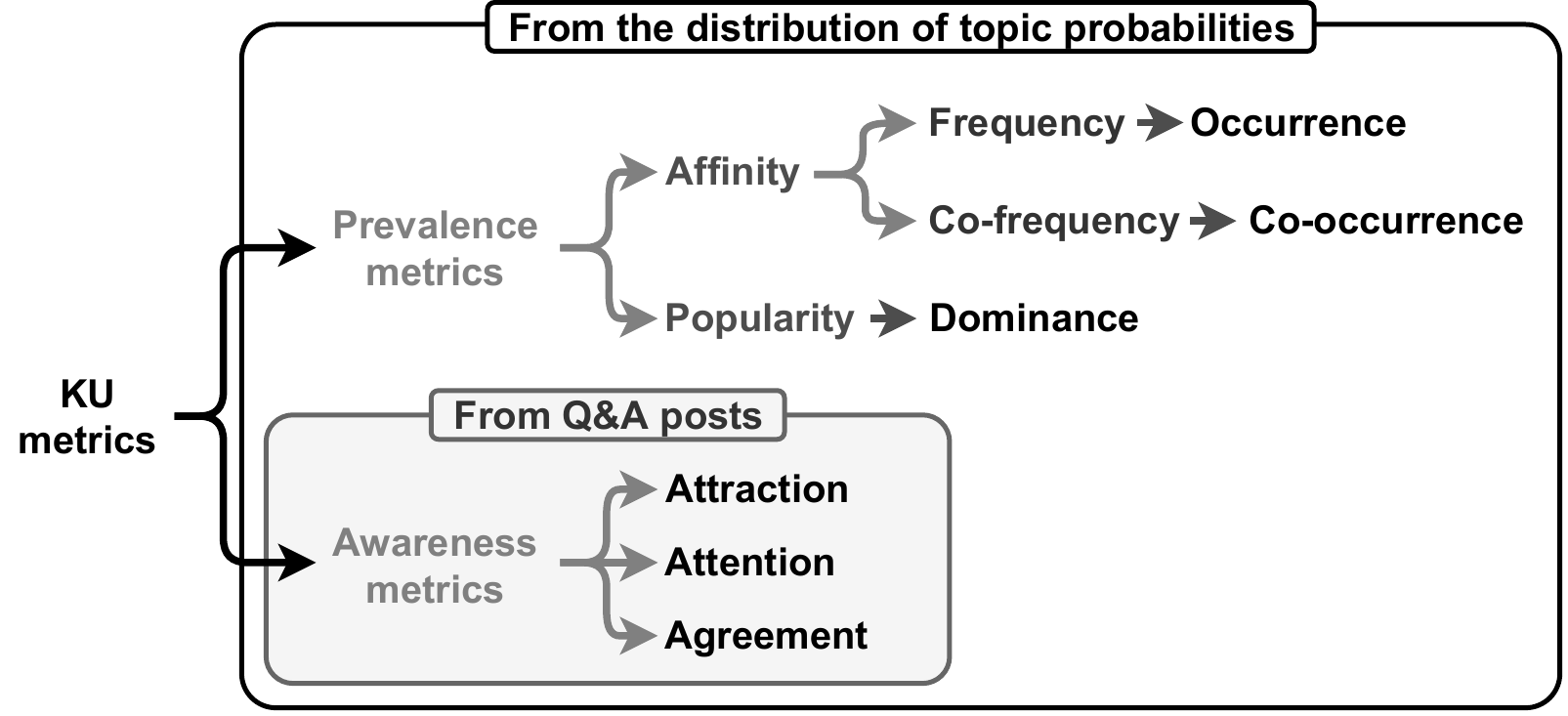}
    \caption{The taxonomy of our KU metrics.}
    \label{fig:ku-metrics}
\end{figure}

The concrete and documentation models represent an estimate of the probability distribution $p(y|x)$ of a certain topic $y$ in a post or a document $x$, given the word types in $x$. We leverage the estimated probabilities of each model to calculate different metrics that measure the topical alignment between the concrete and the documentation models. As we have an one-on-one mapping between KUs and topics, our metrics denote specific characteristics of a KU (e.g., the frequency of a KU) from the measure of properties of the associated topic (e.g., the frequency that the associated topic with the KU occurs in the documents of the corpus). Figure~\ref{fig:ku-metrics} shows the taxonomy of our KU metrics. Two higher-level groups of metrics are the \emph{prevalence} metrics (studied in Sections~\ref{sec:results:subsec:rqone} and~\ref{sec:results:subsec:rqtwo}) and the \emph{awareness} metrics (studied in Section~\ref{sec:results:subsec:rqthree}). Three of the prevalence metrics, namely \textit{occurrence}, \textit{co-occurrence}, and \textit{dominance}, are used as the basis for the definition of other four metrics: \textit{frequency}, which measures how often a topic occurs in a post or document, \textit{co-frequency}, which measures how often two topics co-occur in a post or document, \textit{popularity}, which measures how often a topic has the highest association with a post or document, and \textit{affinity}, which measures the expectation of two topics to occur together in a post or document. Three additional metrics are defined as an awareness metric: \textit{attraction}, which measures the response rate associated with a topic in the Q\&A websites, \textit{attention}, which measures the number of responses to a post in a topic, and \textit{agreement}, which measures the rate of accepted answers associated with a topic. The awareness metrics are calculated from metadata of the Q\&A websites (see Table~\ref{tab:collected-metadata}) and, therefore, are defined only for the concrete model (as we do not have a direct association between a Q\&A post and a document from \rust documentation).

\subsubsection{Calculate prevalence of KUs:}
\label{sec:study-design:subsec:calculate-ku-metrics:subsubsec:prevalence-metrics}

\hfill

\smallskip \noindent \textit{Occurrence:} Q\&A posts and documents can involve related concepts to different KUs. Therefore, we wish to denote the occurrence of a topic in a post or document. A topic $y$ occurs in a post or document $x$ if the topic probability is larger than 50\%:

\begin{align*}
    \textit{occurrence}(y, x) = \begin{cases}
        1, & \text{if}\ p(y|x) > 0.5 \\
        0, & \text{otherwise}
      \end{cases}
\end{align*}

\smallskip \noindent \textit{Co-occurrence:} More than one KU might occur in the same document. Hence, we wish to denote the co-ocurrence of two topics in a document. Two topics $y'$ and $y''$ are co-occurrent if both topics occur in a document $x$:

\begin{align*}
    \textit{co-occurrence}(y', y'', x) = \begin{cases}
        1, & \text{if}\ \textit{occurrence}(y',x) + \textit{occurrence}(y'',x) > 1 \\
        0, & \text{otherwise}
      \end{cases}
\end{align*}

\smallskip \noindent \textit{Dominance:} Although multiple KUs might occur in the same post or document, each of the KUs in a post or document has a different ability to denote the main subject topic. To differentiate a KU that is the main subject of post or document from a KU that is as a supplementary subject, we wish to denote the most representative (a.k.a. dominant) topic from the set of topics in a post or document. The dominant topic of a post or document $x$ is the topic $y$ with the largest probability:

\begin{align*}
    \textit{dominance}(x) = y : p(y|x) = \text{max}(p(y|x)), \forall x \in P
\end{align*}

\smallskip \noindent \textit{Frequency:} A KU can occur more or less frequently in the posts and documents. Hence, we want to measure how often the same topic occurs across the posts or documents. The frequency of a topic $y$ is measured as the number of posts or documents $x$ in which $y$ occurs:

\begin{align*}
    \textit{frequency}(y) = |\{y : \textit{occurrence}(y, x) > 0, \forall x \in P\}|
\end{align*}

\smallskip \noindent \textit{Co-frequency:} Two KUs can occur in conjunton in the same post or document. Hence, we want to measure how often each pair of KUs occurs in a post or document. The co-frequency of a pair of topics $y'$ and $y''$ is the total number of times that these two topics occur in the same post or document $x$.

\begin{align*}
    \textit{co-frequency}(y', y'') = |\{(y', y'') : \textit{coocurrence}(y', y'', x) > 0, \forall x \in P\}|
\end{align*}

\smallskip \noindent \textit{Popularity:} The frequency with which a KU is the main subject topic of a post or document differs depending on the KU. Therefore, we want to measure how often a certain topic is the dominant topic of a post or document. The \emph{popularity} of a topic $y$ is measured as the number of post or documents $x$ in which $y$ is dominant:

\begin{align*}
    \textit{popularity}(y) = |\{y : \textit{dominance}(x) = y, \forall x \in P\}|
\end{align*}

\smallskip \noindent \textit{Affinity:} KUs that often co-occur in the same post or document share a high level of affinity. Although the relatedness of two KUs can be observed from the co-frequency metric, KUs that occur too infrequently in comparison with other KUs are penalized by this metric. We want to measure the \emph{affinity} of a pair of topics normalized by the number of occurrences of both KUs. Affinity is calculated as the number of times that a pair of topics $y'$ and $y''$ co-occur in a post or document divided by the mean frequency of the topics:

\begin{align*}
    \textit{affinity}(y', y'') = \frac{\textit{cofrequency}(y', y'')}{(\textit{frequency}(y') + \textit{frequency}(y'')) /2}
\end{align*}

The affinity metric represents the expectation of two KUs to occur in the same post or document. The metric values range from 0 to 1, with the minimum value representing that the two topics never co-occur and the maximum value representing that whenever one topic occurs in a post or document, the other topic also occurs.

\subsubsection{Calculate awareness of KUs:}
\label{sec:study-design:subsec:calculate-ku-metrics:subsubsec:awareness-metrics}

\hfill

\smallskip \noindent \textit{Attraction:} Associated posts with specific KUs can attract more attention from developers than those associated with others KUs. Q\&A posts that do not receive any comment from developers indicate that developers are not seeking for related information about the KUs that occur in that post. Therefore, to denote the attraction of developers by a KU, we measure the \emph{response rate of a KU} that is dominant in the assciated discussion with a post. Formally, let $a(x)$ be the number of answers of a post $x$. The \emph{attraction} of a topic $y$ is defined by:

\begin{align*}
    \textit{attraction}(y) = \frac{|\{x : \textit{dominance}(x) = y \wedge a(x) > 0, \forall x \in P\}|}{|\{x : \textit{dominance}(x) = y, \forall x \in P\}|}
\end{align*}

The attraction metric ranges from 0 to 1, with the minimum value occurring when none of the Q\&A posts $x$ for which the topic $y$ is dominant have an answer. The maximum value occurs when all posts $x$ for which the topic $y$ is dominant have an answer. The dominant topic is used in this metric because we assume that posts without an answer will be better represented by the dominant topic. If the topic $y$ is never dominant, then attraction is not defined for $y$.

\smallskip \noindent \textit{Attention:} The number of answers that a Q\&A post receives can be understood as the amount of attention that developers draw in that post. Since multiple topics can occur in the same Q\&A post, each topic can contribute with a different weight for attracting attention to that post. The \emph{attention} of a topic $y$ is the average probability $p(y|x)$ weighted by the number of answers in a post $x$:

\begin{align*}
    \textit{attention}(y) = \dfrac{\sum\limits_{x \in P}{\{p(y|x) \times a(x) : \textit{occurrence}(y,x) > 0\}}}{\sum\limits_{x \in P}{\{p(y|x): \textit{occurrence}(y,x) > 0\}}}
\end{align*}

The attention metric represents how much a KU contributes to the total number of answers of the posts for which the KU occurs.

\smallskip \noindent \textit{Agreement:} The degree with which developers agree with the description of the solution of a programming problem can vary. Q\&A posts that have an accepted answer indicate that developers agree with a certain solution to a problem. Therefore, we want to denote the association of a KU with the developers' ability to agree with an accepted answer. Formally, let $a'(x) \neq 0$ if and only if the post $x$ has an accepted answer. We define the \emph{agreement} of a topic $y$ as the proportion of posts $x$ in which $y$ occurs and that have an accepted answer:

\begin{align*}
    \textit{agreement}(y) = \dfrac{|\{x : \textit{occurrence}(y,x) > 0 \wedge a'(x) \neq 0, \forall x \in P\}|}{|\{x: \textit{occurrence}(y,x) > 0, \forall x \in P\}|}
\end{align*}

The agreement metric ranges from 0 to 1, with the minimum value representing that none of the posts that topic $y$ occurs have an accepted answer and the maximum value representing that all posts that topic $y$ occurs have an accepted answer.

\section{Results}
\label{sec:results}

This section describes the motivation, approach, and results of each of our RQs. Sections~\ref{sec:results:subsec:rqone}, \ref{sec:results:subsec:rqtwo}, and \ref{sec:results:subsec:rqthree} discusses RQ1, RQ2, and RQ3, respectively.

\subsection{\rqone}
\label{sec:results:subsec:rqone}


\smallskip \noindent  \textbf{Motivation:} Documenters need to assess how likely is the actual state of the documentation to match the information needs of developers and to reason about the timing for revising the contents of the documentation. In this RQ, we \emph{quantify} the topical alignment between the concrete and the documentation model of \rust by determining how similar is the prevalence of KUs between these two models.

\subsubsection{Analysis of frequency and popularity metrics:}
\label{sec:results:subsec:rqone:subsubsec:frequency-popularity}

\hfill

\smallskip \noindent \textbf{Approach:} For both the concrete and documentation models, we analyze the distribution of the \emph{frequency} and \emph{popularity} metrics along the set of all KUs (see Section~\ref{sec:study-design:subsec:calculate-ku-metrics:subsubsec:prevalence-metrics} for a definition of such metrics). With this analysis, we identify the most frequent and popular KUs in both models and observe the shape of such distributions. To measure the extent to which the frequency and popularity deviate from a normal distribution, we calculate the adjusted Fisher-Pearson standardized moment $G_1$ coefficient of skewness and interpret the absolute value of the coefficient as follows~\cite{weltman:beij:2018}: \textit{almost symetric}, if $-0.5 \leq G_1 \leq 0.5$, \textit{slight skew}, if $0.5 < G_1\leq 1.0$, or \textit{skew}, if $G_1 > 1.0$. We then analyze whether there is a significant correlation between the popularity or the frequency of KUs between the concrete and the documentation models, and whether a significant correlation exists between the two metrics within the same model. From this analysis, we observe the extent to which KUs are equaly prevalent in the two models and whether KUs that are supplementary (as measured by its frequency) are also the main subject (as measured by its popularity) of the individual posts or documents. To measure the correlation, we calculate the Spearman $\rho$ coefficient and interpret its absolute value as follows~\cite{rea:surveydesign:2014}: \textit{negligible}, if $0 \leq |\rho| \leq 0.10$, \textit{weak}, if $0.10 < |\rho| \leq 0.39$, \textit{moderate}, if $0.39 < |\rho| \leq 0.69$, \textit{strong}, if $0.69 < |\rho| \leq 0.89$, or \textit{very strong}, if $0.89 < |\rho| \leq 1$. All hypotheses testing are two-sided and performed under a significance level of $\alpha = 0.05$.



\begin{figure}
    \begin{subfigure}{0.49\textwidth}
        \centering
        \includegraphics[scale=0.19]{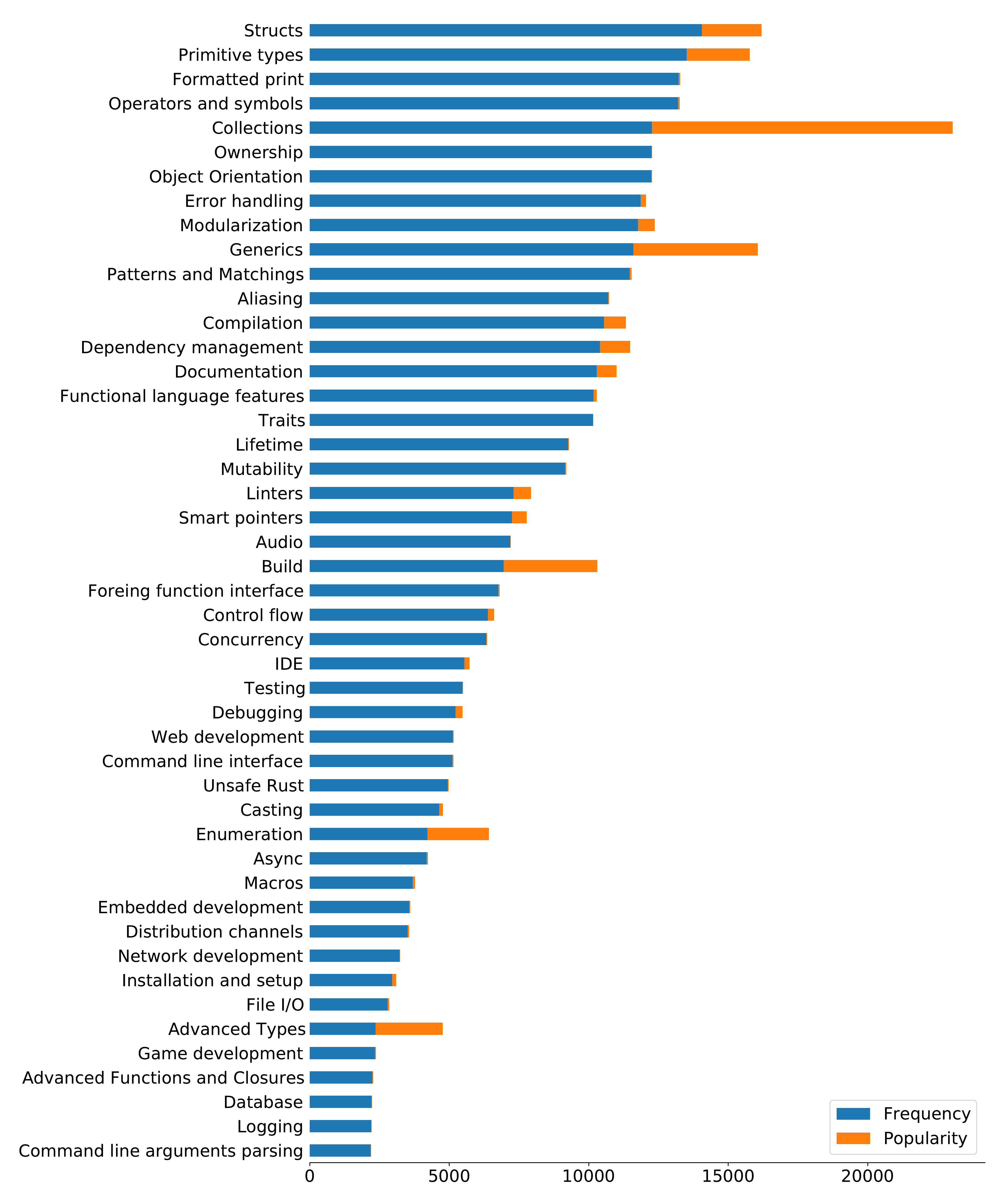}
        \caption{Concrete model}
        \label{fig:topic-prevalence-concrete}
    \end{subfigure}
    \hfill
    \begin{subfigure}{0.49\textwidth}
        \centering
        \includegraphics[scale=0.19]{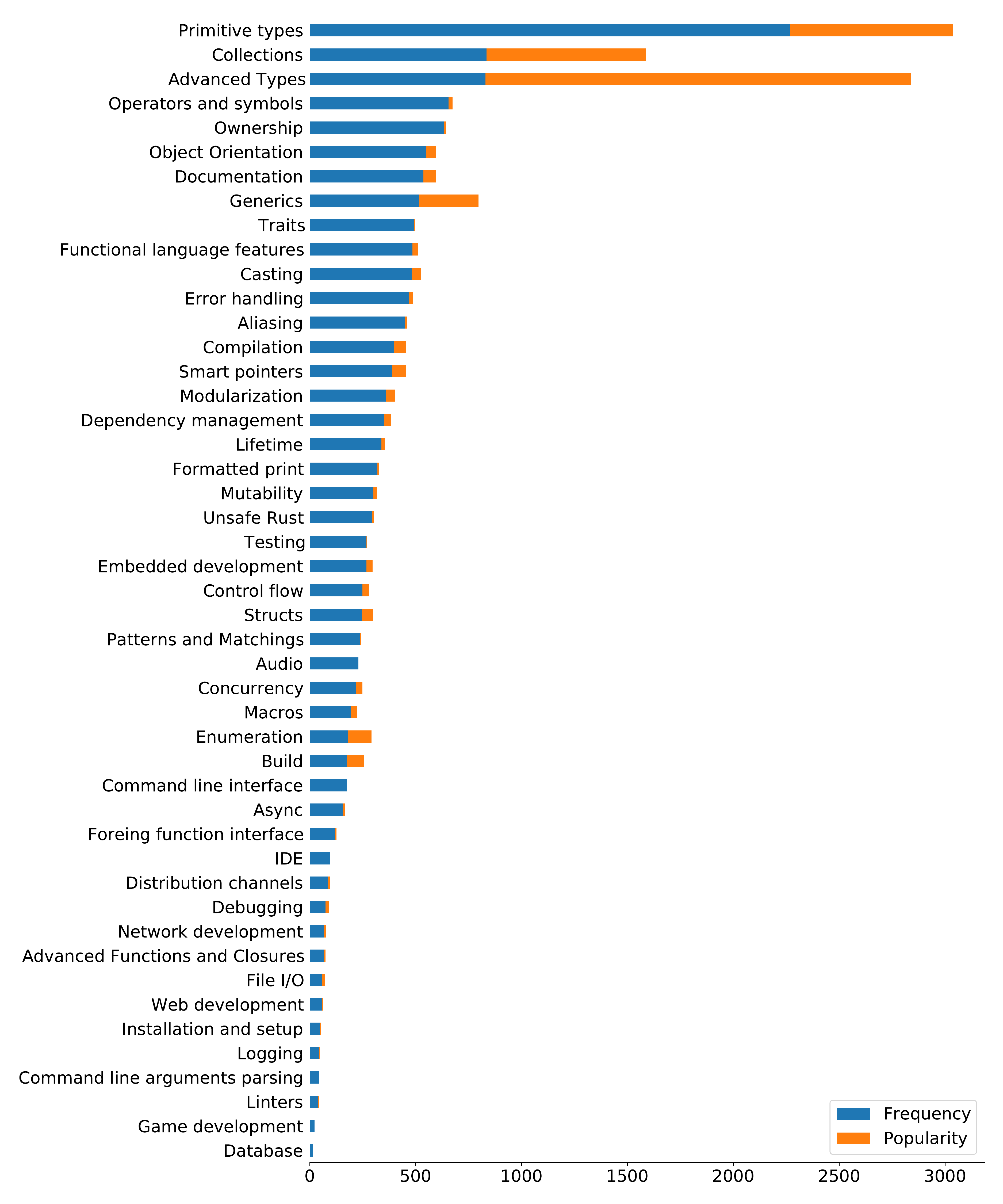}
        \caption{Documentation model}
        \label{fig:topic-prevalence-conceptual}
    \end{subfigure}
    \caption{The frequency and popularity of each KU (sorted by frequency) in the concrete and documentation models of \rust.}
    \label{fig:topic-popularity}
\end{figure}

\smallskip \noindent \textbf{Results:} \textbf{In terms of the rank of frequency and popularity of KUs, there is a strong agreement between the concrete and the documentation models}. Figures~\ref{fig:topic-prevalence-concrete} and \ref{fig:topic-prevalence-conceptual} show both the frequency and popularity of each KU in the concrete and the documentation models, respectively. The frequency and the popularity of the KUs follow a skewed distribution (see Table~\ref{tab:skewness-frequency-popularity}), except for the distribution of the frequency in the documentation model. This observation indicates that few KUs account for most of the individual occurrences and dominances in the documents. Table~\ref{tab:correlation-metrics} shows that there is a strong correlation between the frequency of the KUs of the concrete and the documentation models ($\rho = 0.72$, $p\text{-value} = 1.59 \cdot 10^{-8}$), as well as between the KUs popularity ($\rho = 0.73$, $p\text{-value} = 6.13 \cdot 10^{-9}$). In the concrete model, the frequency of the KUs weakly correlates with the popularity ($\rho = 0.53$, $p\text{-value} = 1.11 \times 10^{-4}$), suggesting the degree to which the same KU occurs either as a supplementary or as the main subject of the posts. In turn, the same correlation is moderate in the documentation model ($\rho = 0.65$, $p\text{-value} = 7.70 \times 10^{-7}$).

\begin{table}
    \caption{The skewness of the distribution of the prevalence metrics in the concrete and conceptual models of \rust documentation.}
    \label{tab:skewness-frequency-popularity}
    \footnotesize
    \begin{tabular}{p{0.15\freewidth}p{0.15\freewidth}p{0.2\freewidth}p{0.12\freewidth}}
        \toprule
        \multirow[t]{2}{*}{\textbf{Model}} & \multirow[t]{2}{*}{\textbf{Metric}} & \multicolumn{2}{l}{\textbf{Skewness}}                        \\* \cmidrule(l){3-4} 
            &   & \textbf{$G_1$ coefficient} & \textbf{Significant?} \\* \cmidrule{1-4}
            Concrete & Frequency & 0.21 (almost symetric) & Yes \\
            & Popularity & 4.36 (skew) & Yes \\ 
            \cmidrule{2-4}
            & Co-frequency & 1.68 (skew) & Yes\\
            & Affinity & 1.46 (skew) & Yes\\ 
            \cmidrule{1-4}
            Documentation & Frequency & 3.64 (skew) & Yes \\
            & Popularity & 4.99 (skew) & Yes \\ 
            \cmidrule{2-4}
            & Co-frequency & 2.78 (skew) & Yes\\
            & Affinity & 1.23 (skew) & Yes\\* 
            \bottomrule
    \end{tabular}
\end{table}


\begin{table}
\caption{The correlation between the same metric for two different models of \rust documentation and the correlation between different metrics for the same model of \rust documentation.}
\label{tab:correlation-metrics}
\footnotesize
\begin{tabular}{p{0.2\freewidth}p{0.28\freewidth}p{0.18\freewidth}p{0.15\freewidth}p{0.12\freewidth}}
    \toprule
    \multirow[t]{4}{*}{\textbf{Metrics}} & \multirow[t]{2}{*}{\textbf{Comparison}} & \multirow[t]{2}{*}{\textbf{Constant}} & \multicolumn{2}{l}{\textbf{Correlation}}                        \\* \cmidrule(l){4-5} 
        & & & \textbf{$\rho$ coefficient} & \textbf{Significant?} \\* \cmidrule{1-5}
        Frequency \& Popularity & \multirow[t]{2}{0.3\freewidth}{Different models, constant metric} & Frequency & 0.72 (strong) & Yes\\
        & & Popularity & 0.73 (strong) & Yes\\ \cmidrule{2-5}
        & \multirow[t]{2}{0.3\freewidth}{Constant model, different metrics} & Concrete model & 0.36 (weak) & Yes\\
        & & Documentation model & 0.65 (moderate) & Yes\\* \cmidrule{1-5}

        Co-frequency \& Affinity & \multirow[t]{2}{0.3\freewidth}{Different models, constant metric}& Co-frequency & 0.79 (strong) & Yes\\
        & & Affinity & 0.69 (strong) & Yes\\ \cmidrule{2-5}
        & \multirow[t]{2}{0.3\freewidth}{Constant model, different metrics}& Concrete model & 0.93 (very strong) & Yes\\
        & & Documentation model & 0.75 (strong) & Yes\\*
        \bottomrule
\end{tabular}
\end{table}

\subsubsection{Analysis of the co-frequency and affinity metrics:}
\label{sec:results:subsec:rqone:subsubsec:cofrequency-affinity}

\hfill

\smallskip \noindent \textbf{Approach:} We first analyze the distribution of the number of KUs that occur in the same post or document of each model. The objective of this analysis is two-fold: first, we want to analyze the extent to which more than one KU is associated with the same post or document, rendering the co-frequency and affinity metrics valid. Second, we want to evaluate how the concrete model agrees with the documentation model in terms of the number of KUs per post or document. We use the one-sided Mann-Whitney procedure to test the null hypothesis that the number of KUs per post of the concrete model is significantly larger than the number of KUs per document of the documentation model ($\alpha = 0.05$).\footnote{The null hypothesis was stated \emph{after} visualizing the distributions of the number of KUs per document in the concrete and documentation models.} In case the null hypothesis is rejected, we calculate the Cliff's Delta $d$ coeficient~\cite{cliff:delta:1996} to measure the effect size of the difference. We interpret the coefficient according to the following thresholds~\cite{romano:amflair:2006}: \textit{negligible}, if $0 \leq |d| < 0.147$, \textit{small}, if $0.147 \leq |d| < 0.330$, \textit{medium}, if $0.330 \leq |d| < 0.474$, or \textit{large}, if $0.474 \leq |d| \leq 1$. As for the co-frequency and the affinity metrics, we first analyze the skewness of the distributions of the two metrics in the two models using the Fisher-Pearson standardized moment $G_1$ coefficient. This analysis shows whether there are pairs of KUs that are more prevalent than others. Then, for both the co-frequency and the affinity metrics of each pair of KUs, we analyze the Spearman $\rho$ rank correlation between the concrete and documentation models, as well as within the same models. The coefficients of skewness and correlation are interpreted according to the thresholds given in Section~\ref{sec:results:subsec:rqone:subsubsec:frequency-popularity}.



\begin{figure}
    \centering
    \includegraphics[scale=0.3]{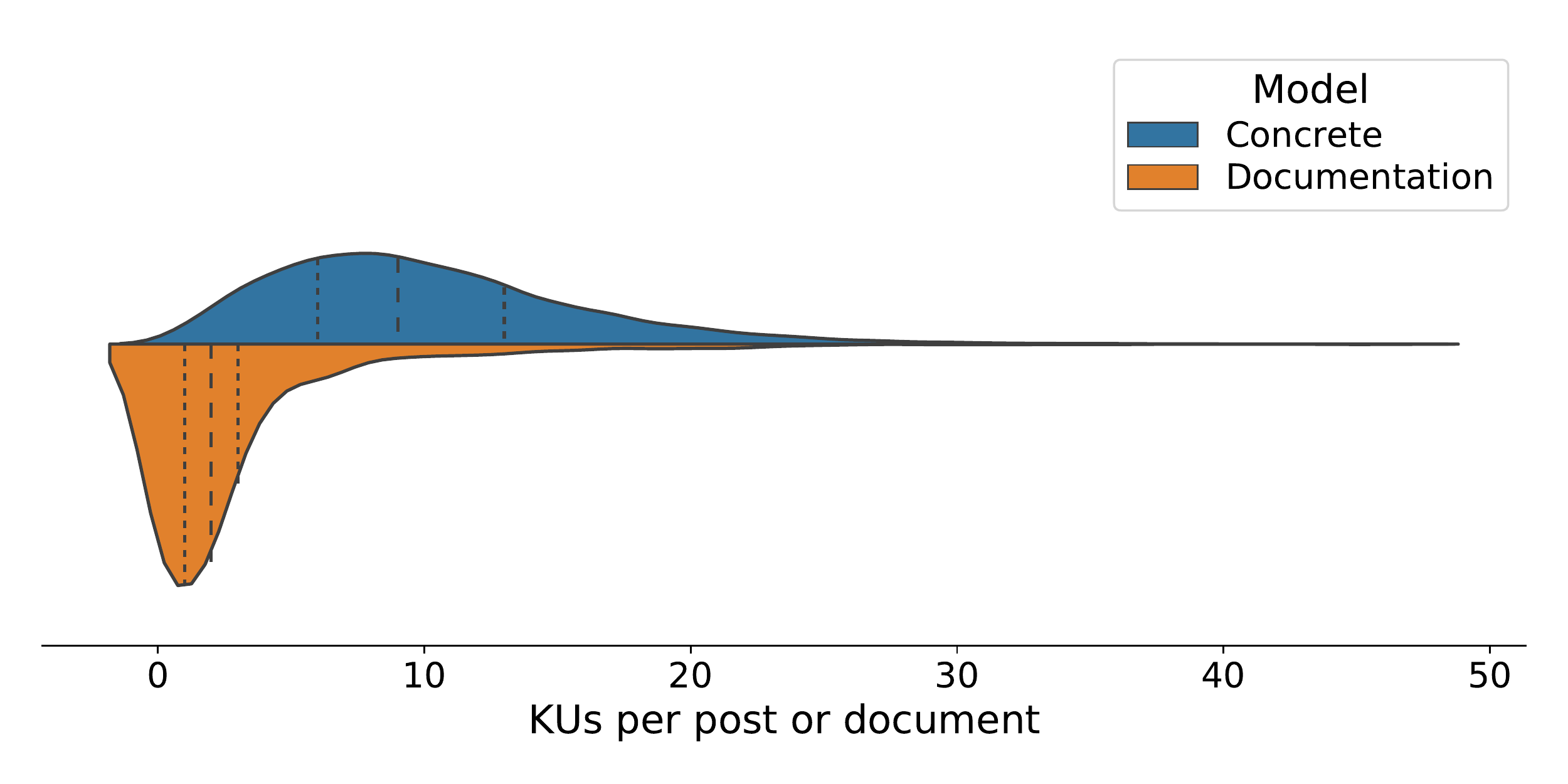}
    \caption{The distribution of the number of KUs that occur in the same document for the concrete and the documentation models of \rust.}
    \label{fig:knowledge-units-per-document}
\end{figure}

\smallskip \noindent \textbf{Results:} \textbf{Multiple KUs occur in association within the same post or document. Still, compared with the documentation model, the concrete model tends to have a significantly larger number of KUs per post}. Figure~\ref{fig:knowledge-units-per-document} shows that the median number of KUs that occur per post is 9 in the concrete model, while the median number of KUs per document in the documentation model is 2. The difference in the number of KUs per post or document between the two models is statistically significant ($p\text{-value} < 0.05$) and the effect size is large ($|d| = 0.748$). As for the co-frequency and affinity metrics, Table~\ref{tab:skewness-frequency-popularity} shows that the distribution of both metrics are skew in both models, suggesting that few pairs of KUs are responsible for most of the co-occurrences in the posts and documents (which also holds valid when we consider the co-occurrences relative to the average frequency of the two KUs). In addition, Table~\ref{tab:correlation-metrics} shows that \textbf{there is a strong correlation between the two models with respect to the rank of the co-frequency and the affinity metrics.} The correlations between the co-frequency and the affinity metrics within the same model are very strong and strong, respectively, showing that the two metrics share similar factors that influence their values within a model.


\begin{tcolorbox}{\textbf{\rqone} \smallskip \\ }
    There is a strong agreement between the concrete and documentation models of \rust in terms of the rank of frequency and popularity metrics. The agreement is also strong for the rank of co-frequency and affinity metrics. Although the co-frequency and affinity metrics have a high correlation within the concrete and documentation models, the same does not hold for the frequency and popularity metrics, for which the correlation within the models are weak and moderate, respectively.
\end{tcolorbox}

\subsection{\rqtwo}
\label{sec:results:subsec:rqtwo}

\smallskip \noindent \textbf{Motivation:} Documenters need to determine which KUs differ between the concrete and documentation models. In particular, documenters can differentiate between KUs that are associated with three different \emph{types of KU alignment}, based on the rank of KUs in the concrete and documentation models with respect to the frequency metric. The three categories of KU alignment are absent, divergent, and convergent. An \emph{absent KU} is a KU that is low ranked in both models. Documenters should be aware of absent KUs as they represent concepts that are rarely documented in both models. A \emph{divergent KU} is a KU that is high ranked in the concrete model but is low ranked in the documentation model. Documenters should consider improving the amount (or enhancing the quality) of information about divergent KUs in the official documentation, as these KUs indicate that there is high demand of information by developers and a scarcity of the same information type in the official documentation. Finally, a \emph{convergent KU} is a KU that is high ranked in both the concrete and the documentation models. Documenters should be mindful of convergent KUs to avoid allocating resources to document redundant information.

\smallskip \noindent \textbf{Approach:} To categorize the KUs into convergent, divergent, and absent, we first divide the rank of KUs into three segments, where the cutoff point of the segments are determined according to the percentiles of the frequency metric. We call a KU as \emph{high ranked KU} if this KU is ranked above the associated cutoff point with the highest ranked segment. Similarly, we call a KU as \emph{low ranked KU} if the KU is ranked below the cutoff point for the lowest ranked segment. As the distribution of the frequency metric has a different shape in the two models (see Section~\ref{sec:results:subsec:rqone:subsubsec:frequency-popularity}), we adopt different cutoff points depending on the model. We tried several combinations of cutoffs and we found that the configuration shown in Table~\ref{tab:cutoff-ranks} yields interpretable and informative results. To aid the interpretability of our analysis of the types of KU alignment, we present the findings using an infographics in which the line color represents the type of KU alignment, while the line thickness represents the rank distance between the same KU on the concrete and documentation models.


\begin{table}
    \caption{Adopted cuttoff points to determine the high and low ranked KUs.}
    \label{tab:cutoff-ranks}
    \footnotesize
    \begin{tabular}{p{0.15\freewidth}p{0.15\freewidth}p{0.15\freewidth}}
    \toprule
    \textbf{Model} & \textbf{High rank cutoff} & \textbf{Low rank cutoff} \\* \cmidrule{1-3}
    Concrete & $50^{th}$ percentile & $25^{th}$ percentile \\
    Documentation & $75^{th}$ percentile & $50^{th}$ percentile \\
    \bottomrule
    \end{tabular}
\end{table}

\smallskip \noindent \textbf{Results:} \textbf{We observe the existence of ten absent KUs} for macros, distribution channels, network development, installation and setup, file I/O, advanced types, game development, advanced functions and closures, database, logging, and command line arguments parsing. In addition to the absent KUs, \textbf{we observed the existence of six divergent KUs} for structs, patterns and matchings, linters, audio, build, and foreign function interface. Finally, \textbf{we also observe the existence of ten convergent KUs} for primitive types, operators and symbols, collections, ownership, object orientation, error handling, generics, documentation, functional language features, and documentation. Figure~\ref{fig:topic-conceptual-alignment} shows the absent, divergent, and convergent KUs in terms of popularity.\footnote{For reference, we also show KUs that were not categorized as absent, divergent, or convergent according to the cuttoff points shown in Table~\ref{tab:cutoff-ranks}.}


\begin{figure}
    \includegraphics[scale=0.25]{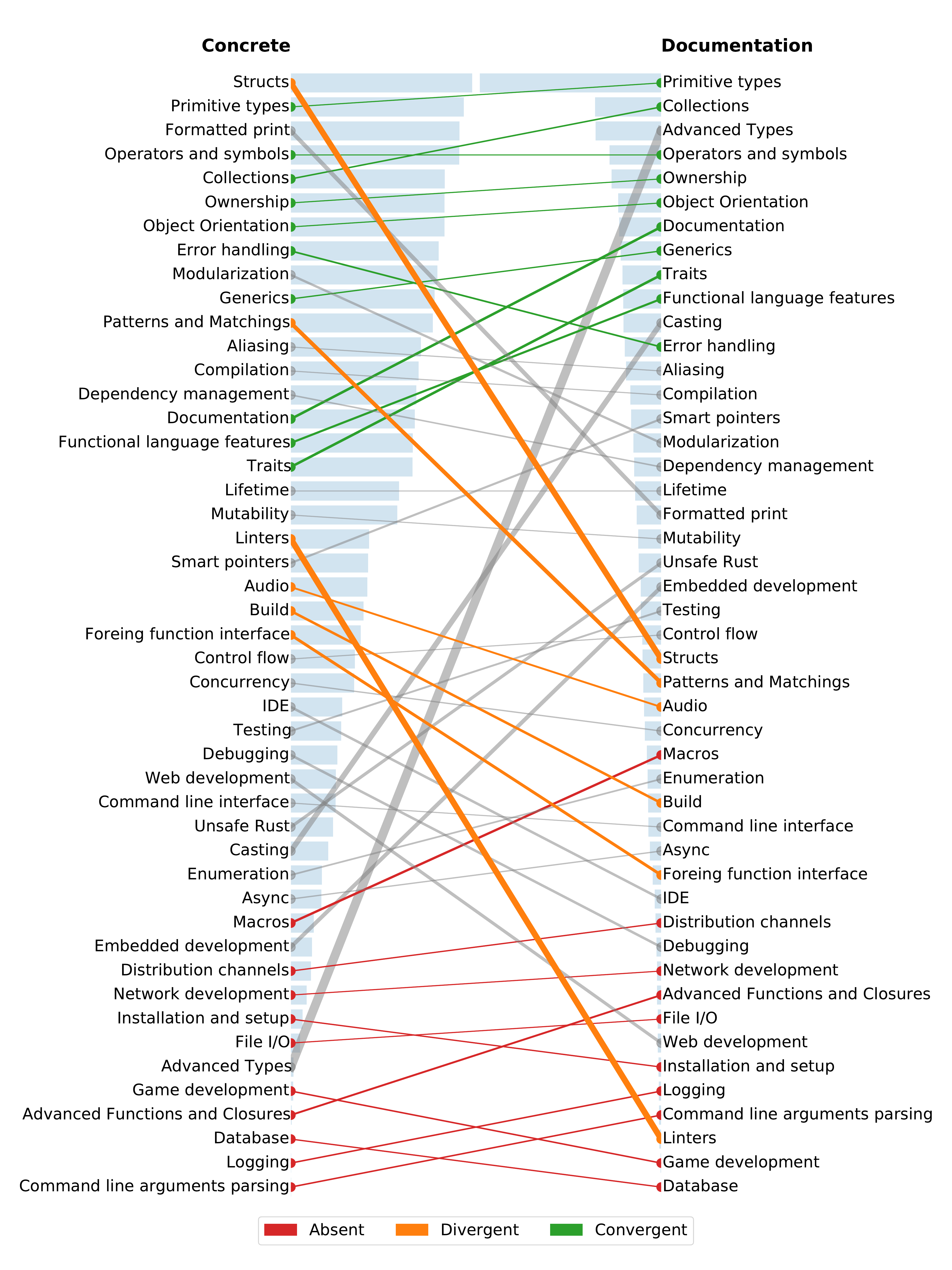}
    \caption{The absent, divergent, and convergent KUs in terms of the frequency metric.}
    \label{fig:topic-conceptual-alignment}
\end{figure}

Table~\ref{tab:category-kus-conceptual-alignment-a} shows the proportion of KUs that pertain to each category grouped by the type of KU alignment (see Table~\ref{tab:ku-category-description} in Section~\ref{sec:study-design:subsec:measuring-concept-align} for a description of the category of KUs). We observe that 50\% of the absent KUs are categorized as a KU for programming niche, suggesting that documenters should consider improving the documentation of these KUs to support the adoption of \rust for those niches. In addition, 40\% of the absent KUs are for the category of language features. Most of the convergent KUs are categorized as KUs for data types (30\%) and language features (60\%), suggesting that documenters do emphasize elementary KUs that occur frequently in the concrete documentation model. Table~\ref{tab:category-kus-conceptual-alignment-b} shows the proportion of absent, divergent, or convergent KUs grouped by the category of the KUs. We can observe that the category of KUs for programming niche has a high proportion of absent KUs (41.6\%). In addition, the category of KUs for data types has a high proportion of convergent KUs (42.8\%). Also noteworthy is the proportion of divergent KUs categorized as language features (33\%), suggesting that the official \rust documentation should consider improving the coverage of this category of KUs.

\begin{table}
    \caption{The relationship between the proportion of each type of KU alignment and the proportion of different categories of KUs.}
    \label{tab:category-kus-conceptual-alignment}
    \footnotesize
    \subfloat[Proportion of KUs in each category of KUs grouped by the type of KU alignment.]{
        \label{tab:category-kus-conceptual-alignment-a}
        \begin{tabular}{p{0.12\freewidth}p{0.18\freewidth}p{0.15\freewidth}}
        \toprule
        \textbf{KU alignment} & \textbf{Category of KUs} & \textbf{Proportion of KUs} \\* \cmidrule{1-3} 
        Absent & Development tooling & 10.0\% \\
        & Language features & 40.0\% \\
        & Programming niche & 50.0\% \\

        Divergent & Data types & 16.7\% \\
        & Development tooling & 33.3\% \\
        & Language features	 & 33.3\% \\
        & Programming niche & 16.7\% \\

        Convergent & Data types & 30.0\% \\
        & Development tooling & 10.0\% \\
        & Language features & 60.0\% \\*
        \bottomrule
    \end{tabular}}
    \quad
    \subfloat[The proportion of absent, divergent, and convergent KUs grouped by the category of KUs.]{
        \label{tab:category-kus-conceptual-alignment-b}
        \begin{tabular}{p{0.18\freewidth}p{0.12\freewidth}p{0.15\freewidth}}
        \toprule
        \textbf{Category of KUs} & \textbf{KU alignment} & \textbf{Proportion of KUs} \\* \cmidrule{1-3} 
        Data types & Divergent & 14.2\% \\
        & Convergent & 42.8\% \\

        Development tooling & Absent & 10.0\% \\
        & Divergent & 20.0\% \\
        & Convergent & 10.0\% \\

        Language features & Absent & 22.0\% \\
        & Divergent & 33.0\% \\
        & Convergent & 11.0\% \\

        Programming niche & Absent & 41.6\% \\
        & Divergent & 8.3\% \\*
        \bottomrule
    \end{tabular}}
\end{table}


\begin{tcolorbox}{\textbf{\rqtwo} \smallskip \\ }
    In addition to six divergent KUs and ten convergent KUs, we identify ten absent KUs in terms of the frequency metric, half of them categorized as KUs for programming niches (e.g., network, game, and database development). Also, one third of the KUs that are categorized as language features are divergent.
\end{tcolorbox}

\subsection{\rqthree}
\label{sec:results:subsec:rqthree}

\smallskip \noindent \textbf{Motivation:} Documenters will benefit from recognizing KUs that need to be prioritized by their documentation efforts. By triangulating awareness metrics from the concrete documentation model with the types of KU alignment, documenters can reason about KUs that should be prioritized in the documentation. In particular, documenters can assess the extent to which developers can find information about convergent, divergent and absent KUs in the form of answers to Q\&A posts and, based on this assessment, decide how to prioritize certain KUs to update the maintained documentation.


\smallskip \noindent \textbf{Approach:} We initially calculate the three awareness metrics for all KUs, from which we obtain three ranked lists of KUs and their respective median attraction, attention, and agreement. We then use the Spearman $\rho$ rank correlation to verify whether there is any significant difference between each pair of awareness metrics that are calculated for the KUs (the estimated $rho$ coefficient is interpreted as described in Section~\ref{sec:results:subsec:rqone:subsubsec:frequency-popularity}). 

Next, we categorize each KU according to their type of KU alignment (see Section~\ref{sec:results:subsec:rqtwo}) and calculate the median of each metric for each of the three types of KU alignment. We then compare the rank of the absent, divergent, and convergent KUs for the same metric. Our comparison consists of using the Mann-Whitney procedure ($\alpha = 0.05$) to perform a post-hoc hypothesis testing of the difference between one type of KU alignment (e.g., convergent KUs) and another type of KU alignment (e.g., divergent KUs) for a given awareness metric. For each of the awareness metrics, we test three hypotheses that compares convergent vs. divergent KUs, convergent vs. absent KUs, and divergent vs. absent KUs. After observing the median of each awareness metric for each type of KU alignment, we decide whether to choose a one- (difference between medians is less than 5\%) or two-sided (difference between medians is greater or equals to 5\%) test and, when a two-sided is chosen, the direction of the test. We correct the multiple hypothesis tests using the Bonferroni procedure~\cite{armstrong:bonferroni:2014} over the groups of equivalent comparisons across the different metrics. For instance, we correct the three tests for the convergent vs. divergent comparison across the attraction, attention, and agreement metrics. For each rejected null hypothesis, we calculate the magnitude of the difference using the Cliff's Delta estimator of effect size, with the same interpretation of the estimates as described in Section~\ref{sec:results:subsec:rqtwo}.

\smallskip \noindent \textbf{Results:} \textbf{Three out of four KUs with the lowest rank for attraction are absent in the \rust documentation, indicating KUs that occur in few documents and often remain unanswered.} Figure~\ref{fig:topic-conceptual-alignment} shows the ranked KUs by their attraction, attention, and agreement metrics. We observe a relatively uniform distribution of values for the same metric across the different KUs. The median attraction, attention, and agreement are, respectively, 82.2\%, 4.66, and 48.0\%. The median value of each metric grouped by type of KU alignment is shown in Table~\ref{tab:comparison-ku-aligment-awareness}. In terms of correlation between the awareness metrics, the only correlated rank of KUs is between the attention and agreement metrics ($\rho = -0.653$, moderate) with the correlation estimate suggesting a decreasing monotonic relationship between the ranks (i.e., while the rank of a KU is high for the attention metric, the rank for the same KU is low for the agreement metric and vice-versa). The correlation between the rank of KUs for the attraction and attention, as well as for the attraction and agreement, is not statistically significant ($\alpha = 0.05$), suggesting that these metrics do not share common factors associated to their values.

\begin{figure}
    \begin{subfigure}{0.49\textwidth}
    \includegraphics[scale=0.17]{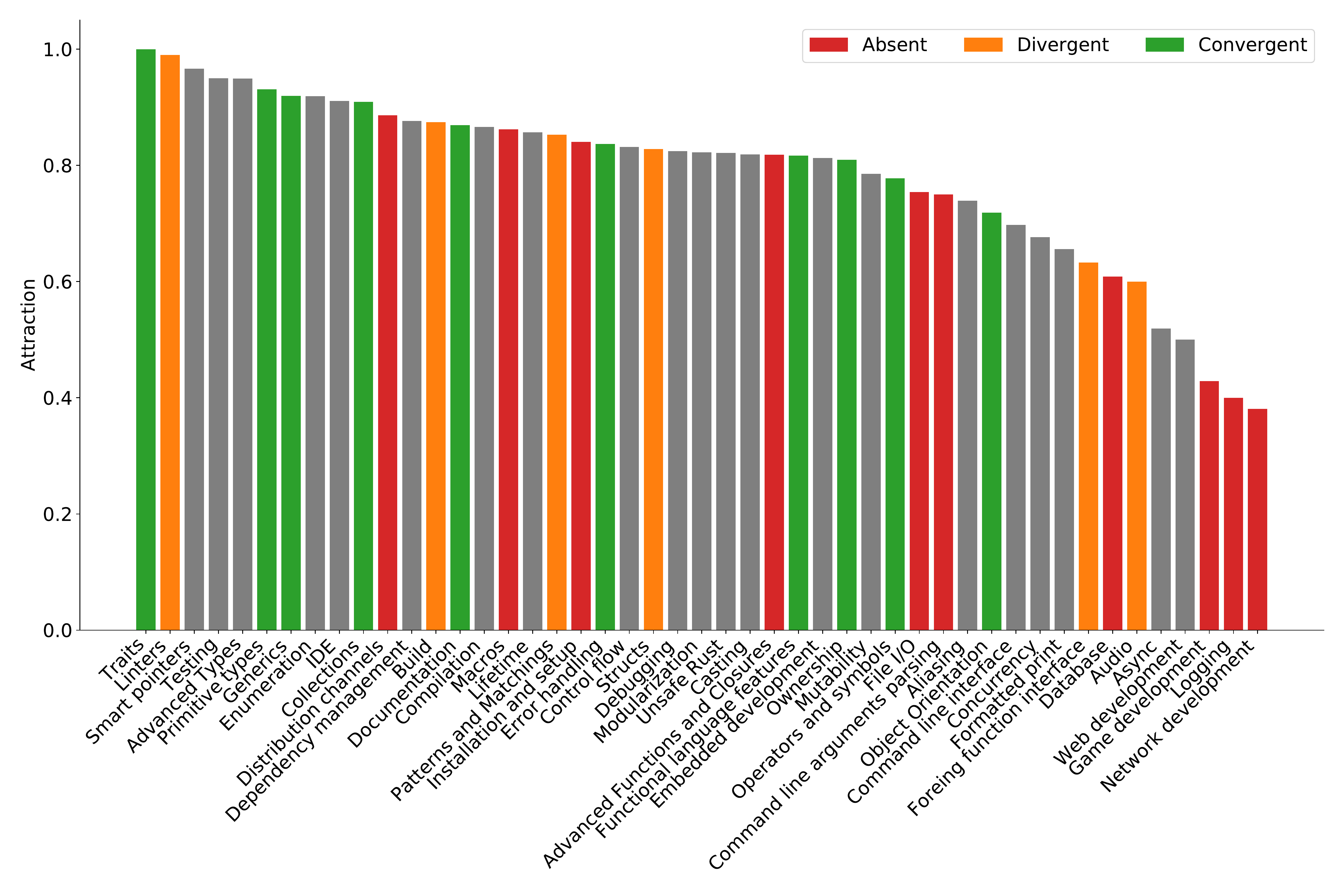}
    \caption{Attraction metric}
    \label{fig:attraction-conceptual-alignment}
    \end{subfigure}
    \begin{subfigure}{0.49\textwidth}
    \includegraphics[scale=0.17]{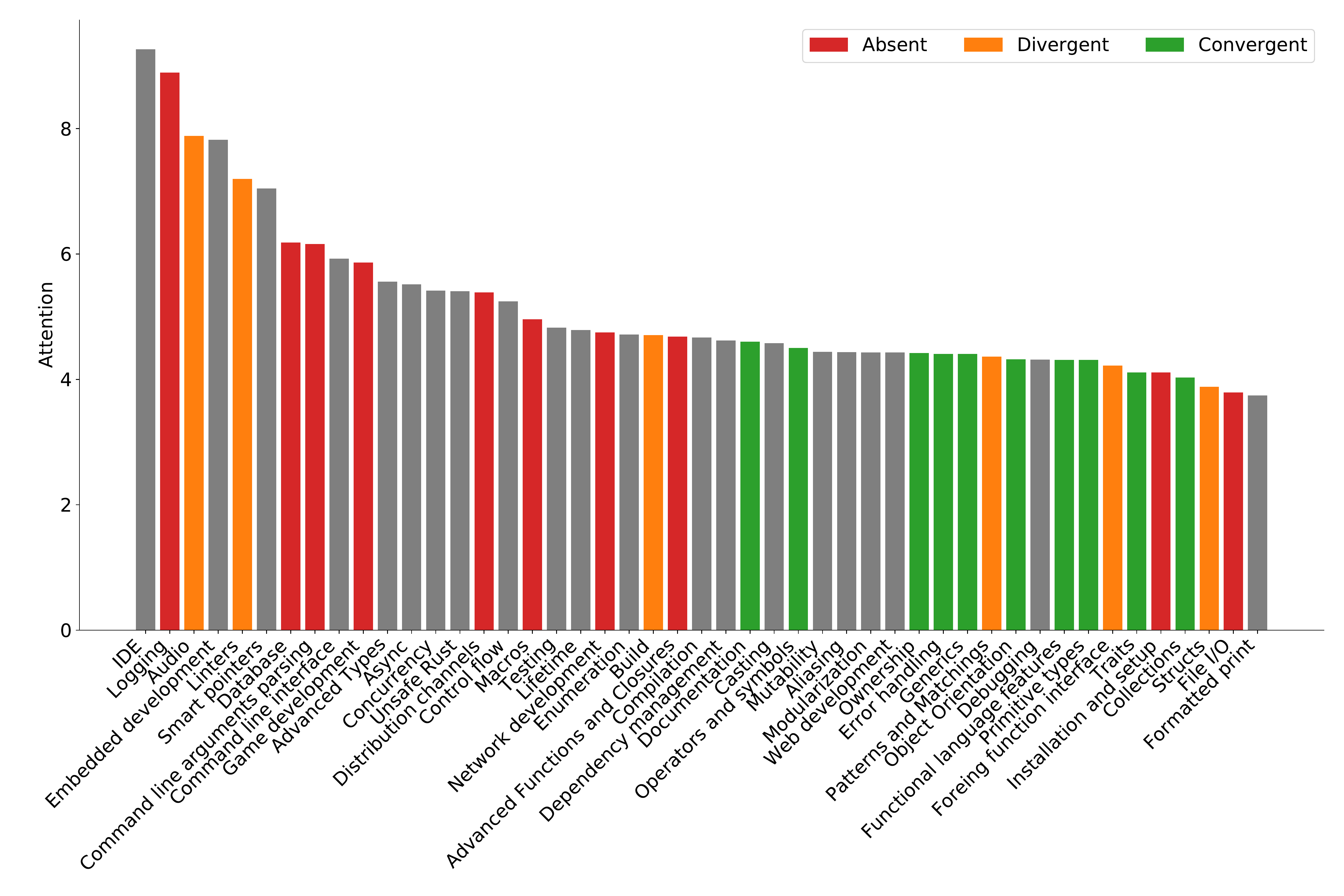}
    \caption{Attention metric}
    \label{fig:attention-conceptual-alignment}
    \end{subfigure}
    \begin{subfigure}{0.49\textwidth}
    \includegraphics[scale=0.17]{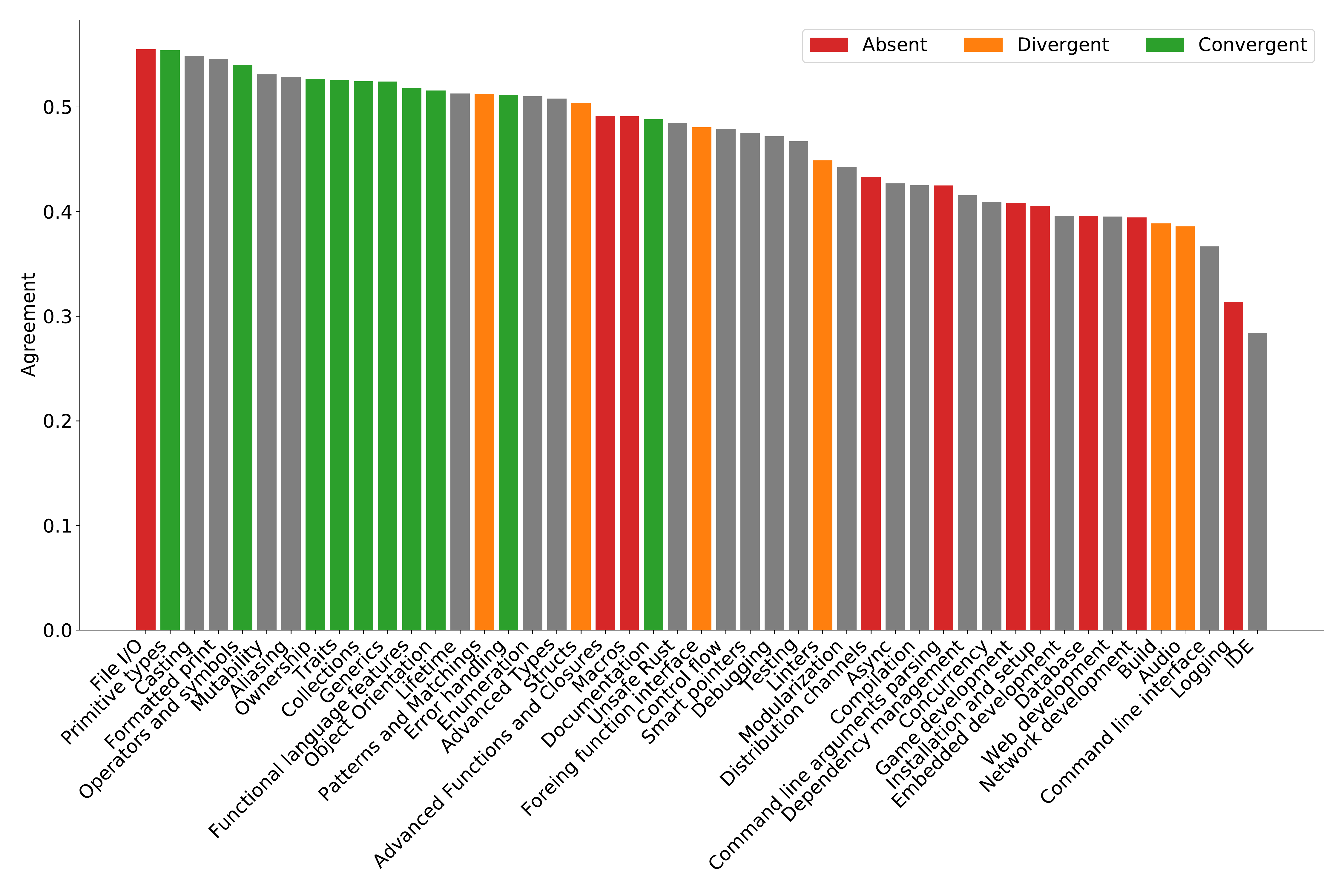}
    \caption{Agreement metric}
    \label{fig:agreement-conceptual-alignment}
    \end{subfigure}
    \caption{KUs ranked by their attraction, attention, and agreement metrics.}
    \label{fig:topic-conceptual-alignment}
\end{figure}

Observing Figure~\ref{fig:attraction-conceptual-alignment}, \textbf{there is no evident pattern that separates absent, divergent, and convergent KUs in the rank of KUs by the attraction metric.} Nonetheless, the rank of KUs by the attention and agreement metrics present clear yet contrary patterns. Figure~\ref{fig:attention-conceptual-alignment} suggests that \textbf{the attention metric tends to be larger for absent and divergent KUs than for convergent KUs}, whereas the difference between absent and divergent KUs is not apparent. In turn, Figure~\ref{fig:agreement-conceptual-alignment} suggests that \textbf{the agreement metric tends to be larger for convergent KUs than for absent and divergent KUs}, while the difference between absent and divergent KUs is also unclear. Table~\ref{tab:comparison-awareness-metrics} shows the results of our hypothesis tests, which partially confirm our initial observation of Figure~\ref{fig:topic-conceptual-alignment}. Particularly for the attraction metric, the difference between the rank of absent, divergent, and convergent KUs is not statistically significant. Also, divergent and absent KUs have no statistically significant differences for any of the awareness metrics. However, the difference between convergent and absent KUs is statistically significant for both the attention and agreement metrics, as well as the difference between convergent and divergent KUs with respect to the agreement metric. All the statistically significant differences have a large effect size.

The observation that the attention of convergent KUs is systematically smaller than that of absent KUs indicates that convergent KUs typically occur in Q\&A posts that receive fewer answers. However, by definition, convergent KUs are also associated with a larger proportion of the documentation. The high amount of associated documents can be related to the fact that most of such KUs are for data types and language features, i.e., they are elementary KUs (see Section~\ref{sec:results:subsec:rqone:subsubsec:frequency-popularity}) that are required by almost every \rust developer and related to many other KUs. In turn, the low amount of answers in Q\&A posts can be related to the fact that information about such elementary KUs occurs with a high frequency in the concrete documentation and, therefore, generate fewer discussions in the form of answers in Q\&A posts. The aforementioned conjectures about the attention of convergent and absent KUs are supported by triangulation with the agreement of those two KU alignments. The agreement for convergent KUs is significantly larger than for absent (and divergent) KUs, suggesting that developers are more likely to accept an answer for a post that is related to a thoroughly documented KU.

\begin{table}
    \caption{The relationship between awareness metrics and types of KU alignment.}
    \label{tab:awareness-ku-alignment}
    \footnotesize
    \subfloat[Median attraction, attention, and agreement metrics for absent, divergent, and convergent KUs.]{
        \label{tab:comparison-ku-aligment-awareness}
        \begin{tabular}{p{0.12\freewidth}p{0.14\freewidth}p{0.07\freewidth}}
        \toprule
        \textbf{KU alignment} & \textbf{Awareness metric} & \textbf{Median} \\* \cmidrule{1-3} 
        Absent & Attraction & 75.0\% \\
        & Attention &  5.17 \\
        & Agreement & 41.0\% \\ \cmidrule{1-3} 

        Divergent & Attraction & 84.0\% \\
        & Attention & 4.53 \\
        & Agreement	 & 46.0\% \\ \cmidrule{1-3} 

        Convergent & Attraction & 85.0\% \\
        & Attention & 4.36 \\
        & Agreement & 52.0\% \\*
        \bottomrule
    \end{tabular}}
    \quad
    \subfloat[Rank-based comparison between absent, divergent, and convergent KUs for attraction, attention, and agreement metrics.]{
    \label{tab:comparison-awareness-metrics}
    \begin{tabular}{p{0.14\freewidth}p{0.18\freewidth}p{0.09\freewidth}p{0.13\freewidth}}
        \toprule
        \textbf{Awareness metric} & \textbf{Null hypothesis ($H_0$)} & \textbf{Reject $H_0$?} & \textbf{Cliff's Delta} ($d$) \\* \cmidrule{1-4} 
            Attraction & Convergent $=$ divergent & No & --- \\
            & Convergent $\leq$ absent & \textbf{Yes} & $0.56$ (large) \\
            & Divergent $\leq$ absent & No & --- \\ \cmidrule{1-4}
            
            Attention & Convergent $\geq$ divergent & No & --- \\
            & Convergent $\geq$ absent & \textbf{Yes} & $-0.62$ (large) \\
            & Divergent $=$ absent & No & --- \\ \cmidrule{1-4}
            
            Agreement & Convergent $\leq$ divergent & \textbf{Yes} & $0.90$ (large) \\
            & Convergent $\leq$ absent & \textbf{Yes} & $0.76$ (large) \\
            & Divergent $\leq$ absent & No & --- \\*
            \bottomrule
    \end{tabular}}
\end{table}

\begin{tcolorbox}{\textbf{\rqthree} \smallskip \\}
    Absent and convergent KUs tend to have opposite patterns of attention and agreement. Absent KUs have high ranked values for the attention metric and low ranked values for the agreement metric. Conversely, convergent KUs have low ranked values for attention and high ranked values for agreement. A statistically significant difference between the ranks of KUs for the divergent and convergent KUs concerning the agreement metric exists. All these differences regarding the ranks of KUs have a large effect size.
\end{tcolorbox}

\section{Discussion}
\label{sec:discussion}

This section discusses the results described in Section~\ref{sec:results}. We draw practical implications for documenters of programming languages and externally validate our results. Our discussion is organized into two sections. In Section~\ref{sec:discussion:subsec:seeking-information-rust}, we discuss the usage of our approach by documenters that want to decide what topics of the documentation should be prioritized when updating the documentation. In Section~\ref{sec:discussion:subsec:seeking-information-rust}, we triangulate our results with external resources to validate our approach to measure the topical alignment of \rust.

\subsection{Prioritization of documentation contents and topical alignment}
\label{sec:discussion:subsec:using-our-approach}

The generated results by applying our approach can be leveraged by documenters to identify KUs that are not well covered by the documentation and decide which of such KUs to prioritize when updating the documentation. For instance, an analysis of the absent KUs suggests documentation topics that should be improved, as absent KUs lack in both the concrete and the documentation models. To decide which absent KUs should be prioritized when incorporating new information into the documentation, documenters can search for absent KUs that often attract responses in Q\&A posts (i.e., high attraction or attention) but still have a low rate of agreement. One example is the KU for distribution channels that, although being highly ranked in terms of attraction, is low ranked in terms of agreement. The same triangulation can be performed for divergent KUs. When deciding which divergent KUs to prioritize, documenters should consider the rank of the awareness metrics associated with those KUs. For example, documenters should consider prioritizing divergent KUs with higher levels of attraction or attention and lower levels of agreement, as information seekers will find it challenging to obtain accepted answers about divergent KUs from Q\&A posts. One example is the KU for build, which is associated with high attention but still exhibits low agreement. In addition, by analyzing the agreement metric of convergent KUs, documenters can spot information that, although widely covered by the official documentation, has a high number of posts that remain unanswered in Q\&A websites. 

When designing instruments for collecting data from the documentation users (e.g., surveys), documenters can leverage our approach to decide with more confidence which questions to emphasize. One of the approaches used by documenters to understand the opportunities to improve a programming language's official documentation is to release surveys to developers. Despite the accurate, detailed, and significant information received through surveys, collecting and analyzing data using this instrument can require high amounts of time and effort. One important lesson learned by applying our approach to assess the topical alignment of \rust documentation is that it is relatively easy to obtain data to build a concrete model of the information needs of developers. After setting up a basic infrastructure for data collection and model training, documenters can efficiently perform iterative analyses to validate and refine hypotheses and complement the results of their survey efforts. For example, consider a closed question of the latest survey released by the \textsf{Go} community that asks ``when using official \textsf{Go} documentation, have you struggled with any of the following''. Among all answers for this question ($n = 2,476$), 31\% selected ``none of the statements apply to me'', showing that documenters will benefit from having a form of iteratively refining research questions. Furthermore, documenters can vary the form of domain knowledge specification (e.g., by considering different KUs or associating different anchor words to the existing KUs), such that different KUs of interest can be incorporated into the documentation models, and new analysis can be performed about more specific or general topics. As another example of the \textsf{Go} survey, the question ``how helpful is official \textsf{Go} documentation for achieving your programming goals in the following areas'' contains a set of seven pre-specified answers. Based on the obtained responses, documenters can use our approach to further understand the documentation of related KUs with the ``least helpful'' areas. More concretely, in the example question of the \textsf{Go} survey, the most popular answer was ``using modules''. Documenters can then hypothesize about a set of modules that developers might be encountering difficulty finding documentation about and add those modules as KUs to a new iterative run of our approach. Nonetheless, our approach is not intended to replace surveys as a form of assessing developers' information needs but, rather, to complement this instrument. In this sense, much of the analyses of the KU metrics can be used to either inform the design of the documentation or to inform the development of other data collection instruments, such as surveys.

In Section~\ref{sec:results:subsec:rqone:subsubsec:cofrequency-affinity}, we found that multiple KUs typically co-occur in the same document, which suggests that documenting information about \rust often requires the compilation of multiple subjects in a single document. We also compared the distribution of the number of KUs that co-occur in the same document between the concrete and documentation models. We found that a large difference exists between the distributions. These observations suggest that documenters should carefully consider how KUs should be combined in the same document to match developers' information needs. A possible solution that documenters can consider regarding the presentation of multiple KUs is cross-linking related KUs in the official documentation. The distribution of the co-frequency and affinity metrics of the concrete model can be used to assess which required information by developers typically involves more than one KU. The co-frequency and affinity metrics can highlight pairs of KUs for which documenters should pay attention to the differences and similarities. For instance, by manually inspecting the ranked list of pairs of KUs with the highest affinity, we observe that the KUs for traits and object orientation have a high affinity. Although this observation is not surprising, given that both KUs share a remarkably high number of related concepts, in fact, traits do not support all features of an object in a strict sense. In particular, the dynamic binding mechanism of traits does not support the concept of inheritance typically found in languages designed as object-oriented (e.g., \textsf{Java}). Therefore, pairs of KUs with high co-frequency and affinity can help documenters to identify KUs for which developers will benefit from examples that promote the comparison between related concepts of one KU to another, as well as KUs whose differences and similarities should be better emphasized in the documentation.


\subsection{External validation}
\label{sec:discussion:subsec:seeking-information-rust}


In Section~\ref{sec:results:subsec:rqone}, we found that the rank of prevalent KUs is similar between the concrete and documentation models. This observation suggests that, with some exceptions, developers seeking for information will find similar distributions of KUs among posts of Q\&A websites and the official documentation. Still, it is valuable to pay attention to the absent and divergent KUs, as they represent topical misalignments. Table~\ref{tab:category-kus-conceptual-alignment-a} shows that the associated KUs with programming niches are at a higher risk of being absent from the \rust documentation. In addition, a close inspection in the ranked KUs by attraction shows that the KU for game development receives fewer answers than other absent KUs. This observation can be externally validated by examining the \rust Game Development Survey released in 2020~\cite{rust:game_survey:2020}. When asked about the negative aspects of using \rust for game development, respondents often point to documentation as one of these aspects, which confirms that developers do have an interest in improving the documentation related to the KU for game development. It is also noteworthy the willingness of the \rust community to promote the adoption of the language for other niches that are associated with absent KUs. For example, a working group for database development (one of the absent KUs) was recently created, and the \rust users already publicly expressed their wish for documentation improvement: ``\textit{I'd love to see some ``database book'' that covers connections, drivers, pools, has examples for every feature and gives overview of ecosystem. Individual project would benefit greatly from better docs}''~\cite{rust:database_wg:2019}. These observations suggest that engaged communities promoting the adoption of \rust for programming niches that are associated with absent KUs should attract documenters. Interestingly, some well documented programming niches (e.g., embedded development, which has a specialized book in the official documentation~\cite{rust:embedded_book:2019}), are not deemed as divergent nor absent, suggesting that these niches are likely well prepared for a potentially growing community of developers.

In the last few years, the \rust community ran the Rust Survey~\cite{rust:rustsurvey:2020} to understand the evolution, challenges, and opportunities of the language. One of the survey questions regards the difficulty of specific topics. Participants have pointed out that lifetime, ownership, macros, trait bounds, and async are the most ``tricky or very difficult'' topics, respectively. We observe that, while the KUs for traits and ownership are perceived as difficult, they are also associated with convergent KUs. A plausible relationship between the perceived difficulty of such KUs and their high frequency on Q\&A websites and the documentation is that information about those KUs might be noisy and often scattered over many documents, making it difficult for developers to pinpoint the information that discusses specificities of these KUs. Therefore, documenters could consider new forms of organizing documentation regarding those KUs, focusing on understanding specific pain points from developers on those KUs, and tackling more specific issues around those KUs in the documentation. Additionally, the KU for macros is both absent and perceived as difficult by developers, raising a flag to documenters regarding the quality and completeness of the documentation around this specific KU.

\section{Threats to validity}
\label{sec:threats-to-validity}

This section discuss the threats to the validity of our study.

\smallskip \noindent \textbf{Selection of \rust related posts from \stkovflw:} As we leverage the tags from \stkovflw to identify \rust related posts, our set of posts can be incomplete (i.e., we might have missed some \rust related posts). The incompleteness stems from the fact that some users might use an unrelated tag to \rust that our selection procedure did not capture. Nonetheless, we expect that this threat is satisfactorily mitigated by expanding the list of intially selected posts using the TET and TST thresholds.

\noindent \textbf{Anchor words definition:} Our set of anchor words were directly derived from the official \rust documentation. Still, the set of anchor words were used to train topic models from data from Q\&A websites. Therefore, some of the adopted terminology can differ between the two data sources. Nonetheless, to mitigate this threat, we use the all known related terms to compose our set of associated anchor words with a KU and validated the terminology with a domain expert.   

\noindent \textbf{KU metrics:} A fundamental principle to the validity of our metrics (i.e., the ability of our metrics to capture the prevalence and awareness of a KU) is the quality of the topic-document assignment of our models. With regards to this factor, during our analyses, we performed sanity checkings to ensure that the topics assigned to each post were in fact corrected. We observe that our topic models can reliably determine the set of topics in a post. However, we also observe that the ability of the dominant topic to distinguish the main subject of a post is smaller. We also verified the ability of our metrics to measure distinct phenomena by running different correlation analyses. Although some degree of correlation exist between our metrics, we can verify that these metrics are able to measure different aspects related to the prevalence and difficulty of a KU.

\noindent \textbf{Definition of absent, divergent, and convergent KUs:} To categorize KU into the three types of KU alignment we adopted a cutoff threshold to separate the rank of KUs into segments. We experimented with many different combinations of cutoff values in search for a configuration that allowed us to obtain interpretable and presentable results. For example, during our experimentation, we searched for a cutoff set up that included at least one of the three types of KU alignment and that included a certain number of KU in each type such that we could discuss the results in the paper. Nonetheless, documenters interested in using our approach can experiment with other cutoff values, depending on how conservatively they want to represent the three types of KU alignment.



\section{Conclusions}
\label{sec:conclusions}

In this paper, we present a machine learning-based approach to study the topical alignment of programming languages documentation, defined as the difference between the concrete developers' information needs and the current state of the documentation. Our approach uses a non-parametric semi-supervised topic model to derive the set of topics from posts of Q\&A websites. Such topic model represents the topics that make up the concrete information needs of developers, and it is used to derive the topics of the programming language documentation. We also propose a set of metrics to measure and evaluate the differences and similarities between the topics of the Q\&A websites and the topics of the programming language documentation. Based on our approach, we perform an empirical study using \rust as a case study.

Our empirical study reveals that, in general, the distribution of topics that make up the concrete information needs of developers share a high degree of similarity with the topics of \rust documentation. Nonetheless, our study highlights topics for which the \rust documentation requires improvement, and indicates missing topics from the documentation that developers are not able to obtain answers from Q\&A websites. More generally, we identified that KUs that occur with less frequency in both the Q\&A websites and the official documentation are often associated with KUs that are categorized as programming niches (e.g., network, game, web, and database development). Those KUs also have a a high number of answers in Q\&A websites but, at the same time, have lower odds of having an accepted answer. In addition, KUs that are categorized as language features (e.g., structs, patterns and matchings, and foreign function interface) often occur in Q\&A websites but not in the official documentation of \rust. These observations are noteworthy, as they suggest KUs that should receive priority from documenters when updating the documentation. Our study contributes to the body of knowledge regarding the documentation of programming languages, and is of particular interest for practitioners that are engaged in actively producing information instead of passively consuming information from the documents collection.




\bibliographystyle{ACM-Reference-Format}
\bibliography{main}

\pagebreak

\appendix

\section{Description of the KUs}
\label{apx:kus-description}

Table~\ref{tab:kus-anchor-words} shows the extracted KUs from the official documentation of \rust.

{\footnotesize
\begin{longtable}{p{0.2\freewidth}p{0.3\freewidth}p{0.5\freewidth}}
    \caption{The derived KUs from \rust's official documentation and their respective anchor words.}\label{tab:kus-anchor-words}\\ %
    \toprule
    \textbf{Category} & \textbf{Knowledge Unit} & \textbf{Anchor words} \\* \midrule
    \endfirsthead
    \endhead
    Data types & Primitive types & type, primitive\_type, scalar\_type, compound\_type, tuple, array, scalar, slice \\
    & Structs & struct, named\_field, name\_field, structure \\
    & Enumeration & enum, enumeration, variant \\
    & Collections & common\_collections, common\_collection, collections, collection, vector, hashmap, hash\_map, append, index, std\_collection, stdcollection, std\_collections, stdcollections, vec, hashset, hash\_set \\
    & Generics & generic, generic\_type, concrete, abstract, monomorphization, bound, phantom\_type \\
    & Smart pointers & smart\_pointer, reference\_counting, deref, deref\_trait, drop\_trait, interior\_mutability, reference\_cycles, cons\_list, con\_list, cons\_function, con\_function, reference\_cycle, recursive\_type, dereference\_operator \\
    & Advanced Types & newtype, newtype\_pattern, type\_alias, never\_type, dynamically\_sized\_type, dst, unsized\_types, dynamic\_size\_type, dynamic\_size, unsize\_type, unit\_type \\
    Development tooling & IDE & rls, rust\_analyzer, ide, visual\_studio \\
    & Linters & rustfmt, rustfix, clippy, linter, lint, rust-clippy \\
    & Installation and setup & installation, install, hello\_world, rustup, rustup.rs \\
    & Modularization & module, reuse, external\_code, private, public, visibility, file\_hierarchy, file\_hierarch, reexport, workspace \\
    & Testing & test, automated\_test, automate\_test, assert, unit\_test, integration\_test, cargo\_test \\
    & Dependency management & release, release\_profile, library, crates\_io, yank, cargo\_install, lockfile, dependency, update\_dependency, cargo\_publish, semver \\
    & Compilation & rustc, crate, compilation\_unit, crate\_file, compiler, compilation, flag, compiler\_flag, compile\_flag, compiler\_option, compile\_option, cargo\_rustc, crate\_type, link\_crate, abi, application\_binary\_interface \\
    & Build & build, cargo\_toml, build\_script, build\_rs, cargo\_build \\
    & Documentation & rustdoc, cargo\_doc, cargodoc, doctest, document\_api, rfc1574, rfc1946, pass, passes \\
    & Debugging & debug, std\_fmt, debug\_info, debug\_symbol, debugger, rustdt, msvc\_abi, rust\_gdb, gdb, rust\_lldb \\
    Language features & Ownership & ownership, stack, heap, reference, borrow, move, copy, partial\_move, drop, forget \\
    & Error handling & error\_handling, exception, recoverable\_error, unrecoverable\_error, unrecover\_error, panic, abort, unwind, result, option \\
    & Functional language features & functional\_programming, functional\_style, closure, anonymous, anonymous\_type, anonymous\_function, type\_anonymity, type\_anonym, iterator\_trait, next\_method, hof, high\_order\_function \\
    & Object Orientation & object\_oriented, object\_orientation, oo, oop, object, inheritance, trait \\
    & Patterns and Matchings & pattern\_matching, pattern\_match, match, if\_let, while\_let, matches \\
    & Unsafe Rust & unsafe, unsafety, memory\_unsafety, unsafe\_trait, raw\_pointer, transmute \\
    & Traits & advanced\_trait, associated\_types, operator\_overload, default\_type\_parameter, trait\_implementation, implement\_trait, supertrait \\
    & Advanced Functions and Closures & function\_pointer, function\_point, return\_closure, returning\_closure, hrtb, higher\_rank\_trait\_bound \\
    & Macros & macro, declarative\_macro, procedural\_macro, metaprogramming, derive\_macro \\
    & Operators and symbols & operators, symbols, literal, constant, const \\
    & Distribution channels & nightly, night, distribution\_channel, distribut\_channel, night\_rust, train\_schedule, rust\_release, next\_train, release\_train, feature\_flag, install\_night, unstable\_feature, nightly\_release, night\_release \\
    & Mutability & variable\_binding, mutable, mutability, immutable, mut, mut\_modifier, freeze, frozen \\
    & Lifetime & variable\_shadow, shadow, variable\_binding, block, scope, raii, lifetime, drop, scope, out\_scope, lifetime\_annotation \\
    & Casting & cast, casting, type\_conversion, type\_inference, from\_trait, into\_trait, tryfrom, tryinto, tostring, fromstr \\
    & Aliasing & alias, type\_alias \\
    & Control flow & condition, boolean\_condition, boolean\_condit, control\_flow, decision, loop, break, continue, std\_iter, named\_loop \\
    & File I/O & file\_read, file\_write, file\_method, read\_only, readonly, file\_descriptor, write\_only, writeonly, line\_file, std\_io, open\_options \\
    & Foreing function interface & ffi, foreign\_function\_interface, foreign\_interface, foreign\_function, foreign\_library, snappy, libc, wrap, foreign\_global, no\_mangle, bindgen \\
    Programming niche & Concurrency & concurrency, concurrent, race\_condition, deadlock, multithread, thread, process, spawn, message\_pass, producer, consumer, mutex, mutexes, std\_sync, std\_thread, rayon \\
    & Game development & game, roguelike, sokoban, game\_engine, windowing, physics, rendering, fmod, openal, ggez \\
    & Web development & web, web\_development, web\_api, rocket, actix\_web, diesel, sqlx, oauth, cookie, oauth2, html, css, webassembly, http, http\_request \\
    & Network development & openssl, network, network\_layer, dns, tcp\_ip, ip, tcp, udp, std\_net, ftp, smtp, cryptography, encryption, certificate \\
    & Embedded development & embed, embedded, microcontroller, qemu, interrupt, spi, uart, rs232, usb, i2c, ttl \\
    & Database & database, sqlx, db, mongodb, sqlite, postgres, mysql, sql, driver, orms, orm, elasticsearch \\
    & Logging & logging, log, console\_logger, env\_logger, log4rs, fern, syslog, log \\
    & Audio & audio, midi, music, audio\_decoder, audio\_decode, audio\_encoder, audio\_encode, libsoundio \\
    & Command line interface & command\_line, argument\_parse, ansi\_terminal, ansi\_term, terminal, ansi, termcolor \\
    & Formatted print & format\_print, formatted\_print, formating, format, print, formatting\_trait, format\_trait \\
    & Async & tokio, async\_std, mio, future \\
    & Command line arguments parsing & clap, structopt, std\_env\_arg \\* \bottomrule
\end{longtable}}
\section{Initial tag set}
\label{apx:initial-tag-set}

Table~\ref{tab:initial-set-rust-tags} shows the initial set of \rust related tags.

\begin{table}[H]
    \caption{Our initial set of \rust related tags.}
    \label{tab:initial-set-rust-tags}
    \footnotesize
    \begin{tabular}{p{0.2\freewidth}p{0.2\freewidth}p{0.2\freewidth}}
    \toprule
    \multirow{2}{*}{\textbf{Category}} & \multicolumn{2}{l}{\textbf{Tag}}                                           \\ \cmidrule(l){2-3} 
                                       & \textbf{Derived from Stack Overflow} & \textbf{Derived from documentation} \\ \cmidrule{1-3}
    General                            & ``rust''                             & --                                  \\
    Language feature                   & ``rust-macros''                      & ``borrow''                          \\
                                       & ``rust-proc-macros''                 & ``borrow-checker''                  \\
                                       & ``rust-obsolete''                    & ``supertrait''                      \\
                                       & ``rust-crates''                      & ``unsafe''                          \\
                                       & --                                   & ``refcell''                         \\
                                       & --                                   & ``collections''                     \\
                                       & --                                   & ``crate''                           \\
    Development tool                   & ``rust-cargo''                       & ``rustc''                           \\
                                       & ``rust-tokio''                       & ``rustup''                          \\
                                       & ``rust-diesel'                       & ``rustdoc''                         \\
                                       & ``rust-actix''                       & ``rustfix''                         \\
                                       & ``rust-rocket'                       & ``clippy''                          \\
                                       & ``rust-warp''                        & ``rls''                             \\
                                       & ``rust-chrono''                      & --                                  \\
                                       & ``rust-piston''                      & --                                  \\ \bottomrule
    \end{tabular}%
    \end{table}
\section{Derived topics}
\label{apx:topics}

Table~\ref{tab:topics-so} shows the top 20 words of the semi-supervised topic model associated with the concrete model.

{\footnotesize
\begin{longtable}{p{0.2\freewidth}p{0.8\freewidth}}
    \caption{The top 20 words of each derived topic of the concrete model.}\label{tab:topics-so} \\%
    \toprule
    \footnotesize
    \textbf{Topic (associated KU)}        & \textbf{Word types}  \\ \midrule
    Advanced Types & unstable, newtype, specialization, dst, type\_alias, dynamically\_sized\_type, unsized\_type, unit\_type, unboxed\_closures, fn\_traits, const\_fn, unsize, type\_alias\_impl\_trait, never\_type, newtype\_pattern, type\_ascription, associated\_type\_defaults, dynamic\_size, specialized, newtypes \\
Collections & vec, vector, iter, map, hashmap, element, item, collection, std\_collection, index, string, hashset, collect, hash\_map, push, iterator, len, into\_iter, append, usize \\
Enumeration & enum, variant, enums, enumeration, myenum, union, discriminant, tag, untagged, e0015, nil, exhaustive, tagged, strum, variant1, strum\_macros, variant2, uninhabited, discriminator, boolean \\
Generics & trait, type, generic, parameter, bound, concrete, generic\_type, abstract, note, sized, monomorphization, required, requirement, ops, constraint, known, annotation, inference, mismatch, accept \\
Primitive types & value, type, element, vector, slice, array, tuple, scalar, primitive\_type, length, integer, indexing, tuples, val, f64, pair, f32, arr, contiguous, elem \\
Smart pointers & deref, interior\_mutability, derefmut, case, smart\_pointer, actually, mean, memory, doe, reason, example, different, instead, data, reference\_counting, possible, kind, understand, fact, place \\
Structs & struct, structs, structure, impl, self, fn, pub, field, deserialize, foo, phantomdata, member, serialize, mystruct, bool, eq, name\_field, constructor, named\_field, u32 \\
Build & build, cargo, cargo\_toml, cargo\_build, build\_script, project, toml, github, registry, building, git, 1ecc6299db9ec823, built, manifest, edition, latest, profile, builder, ci, cdylib \\
Compilation & build, cargo, rustc, compilation, compiled, flag, compile, crate, library, abi, crate\_type, cargo\_rustc, compiler, compiling, v0, aborting, compiler\_flag, previous, information, verbose \\
Debugging & debug, std\_fmt, gdb, debugger, derive, debug\_info, partialeq, fmt, breakpoints, debuginfo, formatter, rustdt, unoptimized, display, rust\_gdb, debug\_symbol, serde, codelldb, debugging, openocd \\
Dependency management & library, cargo, dependency, release, package, crate\_io, semver, standard, version, cargo\_install, author, lockfile, yank, party, publish, compatibility, update\_dependency, published, utils, publishing \\
Documentation & test, pas, passed, passing, pass, doctest, cargo\_doc, run\_test, utilizing, new\_string, testresult, arg3, arg2, mycallback, redacted, den, d4h0, spice, fluff, alleviates \\
IDE & ide, vscode, rls, intellij, visual\_studio, thanks, lot, right, maybe, help, post, yes, thank, language, issue, people, try, great, guess, thought \\
Installation and setup & install, installed, hello\_world, rustup, installation, x86\_64, gnu, usr, toolchain, toolchains, rustlib, unknown, ld, libstd, pc, installing, libcore, cc, msvc, ubuntu \\
Linters & closed, topic, reply, day, automatically, wa, allowed, longer, clippy, lint, new, invite, rustfmt, comment, open, question, hello, linting, linter, alice \\
Modularization & crate, module, import, public, private, workspace, exported, visibility, submodule, submodules, publicly, privacy, mod, reexport, external, export, unresolved, importing, imported, external\_code \\
Testing & test, assert, unit\_test, cargo\_test, it\_works, integration\_test, assert\_eq, my\_test, cfg, testing, should\_panic, test\_foo, super, bench, some\_test, assertion, bencher, integration, measured, suite \\
Advanced Functions and Closures & function\_pointer, hrtb, return\_closure, returning\_closure, higher\_rank\_trait\_bound, function\_point, call\_it, add\_closure, higher\_order, amet, dolor, f8, cortex\_m\_semihosting, 0x14, 0x28, 0x00, stm32f4, toabc, fnset, vec\_closure \\
Aliasing & trait, type, parameter, define, alias, type\_alias, defined, associated, definition, following, specify, named, defining, constrained, defines, rh, mul, inherent, unconstrained, ranked \\
Casting & convert, cast, casting, fromstr, type\_inference, tryfrom, tostring, tryinto, type\_conversion, conversion, converting, converted, try\_from, try\_into, tryfromsliceerror, i8, u16, digit, e0605, convertible \\
Control flow & loop, break, continue, condition, decision, control\_flow, iteration, event, controlflow, looping, infinite, skip, std\_iter, quit, subsequent, continuously, met, panic\_handler, iterative, panicinfo \\
Distribution channels & nightly, stable, unstable\_feature, night, nightlies, feature\_flag, feature, rust\_release, nightly\_release, beta, build\_run, experimental, stabilized, rust1, multirust, stability, intrinsic, gate, enabling, stabilization \\
Error handling & result, error, err, option, panic, panicking, abort, exception, error\_handling, unwind, return, catch\_unwind, std, unwrap, src, message, failure, invalid, success, fails \\
File I/O & std\_io, file\_read, file\_descriptor, io, readonly, line\_file, stdin, read, read\_line, file\_write, bufreader, buffer, bufread, reader, byte, expect, buf, stdout, input, line \\
Foreing function interface & extern, wrap, ffi, libc, no\_mangle, c\_void, bindgen, macro\_use, stdcall, c\_double, foreign\_function, link\_name, c\_int, serde\_derive, snappy, typedef, no\_std, rustc\_serialize, c\_str, winapi \\
Functional language features & lifetime, parameter, closure, outlive, function, anonymous, infer, argument, fnmut, conflicting, anonymous\_type, body, fnonce, functional\_programming, anonymous\_function, captured, capture, iterator\_trait, callback, inferred \\
Lifetime & lifetime, reference, dropped, drop, scope, variable, block, live, lifetime\_annotation, long, raii, valid, life, shadow, e0597, static, e0495, temporary, explicit, variable\_binding \\
Macros & macro, procedural, procedural\_macro, macro\_rules, derive\_macro, ident, expr, declarative, declarative\_macro, tt, metaprogramming, expansion, ty, proc, originates, token, tokenstream, expand, generate, macro\_export \\
Mutability & borrow, reference, mutable, mut, immutable, mutate, mutability, checker, borrows, occurs, mutably, refcell, ref, cell, e0502, borrow\_mut, e0499, modify, mutation, content \\
Object Orientation & trait, implement, generic, method, implementing, object, implemented, inheritance, implementation, oop, composition, box, oo, boxed, object\_oriented, polymorphism, partialord, ord, equality, delegate \\
Operators and symbols & function, type, const, array, expression, operator, constant, literal, signature, expected, i32, returning, mismatched, e0308, syntax, declare, opaque, expects, overloading, overload \\
Ownership & reference, borrow, mutable, ownership, value, owned, borrowing, pointer, moved, heap, copy, closure, stack, drop, borrowed, forget, clone, allocated, moving, owns \\
Patterns and Matchings & match, ok, err, let, pattern\_matching, pattern\_match, str, parse, matching, false, errorkind, from\_str, eprintln, arm, msg, as\_str, unreachable, matched, branch, parsing \\
Traits & trait, implement\_trait, associated\_type, trait\_implementation, e0277, supertrait, satisfied, dyn, marker, subtrait, dispatch, impls, blanket, mytrait, e0599, unsized, vtable, e0119, implementors, default\_type\_parameter \\
Unsafe Rust & unsafe, safe, transmute, raw\_pointer, unsafety, from\_raw\_parts\_mut, ptr, memory\_unsafety, mem, assume\_init, soundly, as\_ptr, safety, raw, undefined, as\_mut\_ptr, ub, repr, c\_char, from\_raw\_parts \\
Async & future, async, tokio, mio, await, async\_std, stream, poll, executor, map\_err, ready, asynchronous, and\_then, block\_on, pin, reactor, fut, streamext, tokio\_core, polled \\
Audio & think, thing, make, really, ha, like, good, time, sure, need, audio, know, bit, point, way, probably, better, look, used, problem \\
Command line arguments parsing & clap, subcommand, structopt, subcommands, from\_os\_str, argmatches, rdi, rax, mov, appsettings, rsi, rsp, eax, required\_unless, rcx, rdx, load\_yaml, retq, movq, ret \\
Command line interface & terminal, command\_line, target, file, linux, run, command, running, ansi, bin, window, directory, linker, path, program, exit, failed, link, linking, script \\
Concurrency & thread, spawn, channel, mutex, std\_sync, blocking, std\_thread, task, threaded, process, concurrent, rayon, spawning, concurrently, concurrency, multithreaded, deadlock, mutexes, producer, send \\
Database & database, db, postgres, driver, sql, mysql, sqlite, postgresql, mongodb, sqlx, orm, elasticsearch, query, table, conn, id, schema, select, queryable, row \\
Embedded development & embedded, embed, interrupt, performance, faster, optimization, stm32, microcontroller, qemu, fast, usb, cpu, number, size, speed, slower, large, slow, small, firmware \\
Formatted print & println, print, format, main, to\_string, output, printing, printed, formatting, push\_str, formatted, num, formatting\_trait, format\_print, gen\_range, format\_trait, letter, printer, 02x, uncomment \\
Game development & game, rendering, physic, ggez, windowing, game\_engine, draw, image, width, height, render, graphic, pixel, texture, opengl, gaming, roguelike, color, player, screen \\
Logging & log, logging, env\_logger, levelfilter, log4rs, set\_logger, syslog, simple\_logger, loglevelfilter, setloggererror, fern, logger, console, appenders, production, rust\_log, essential, pipeline, infrastructure, rolling \\
Network development & server, network, tcp, ip, openssl, std\_net, udp, certificate, encryption, connection, socket, dns, tcpstream, ftp, net, bind, telnet, connect, smtp, tcplistener \\
Web development & http, server, web, html, hyper, actix, reqwest, diesel, rocket, actix\_web, webassembly, http\_request, request, cookie, client, cs, com, response, org, lang \\ \bottomrule
\end{longtable}}

\end{document}